\documentclass[a4paper]{jpconf}
\pdfoutput=1

\usepackage{xspace}
\usepackage{graphicx}
\usepackage[font=small,labelfont=bf,belowskip=-10pt]{caption}

\newcommand {\pT}        {\ensuremath{p_\mathrm{T}}\xspace}
\newcommand {\vn}	{\ensuremath{v_\mathrm{n}}\xspace}
\newcommand {\meanpT}    {\ensuremath{\langle p_{\mathrm{T}} \kern-0.1em\rangle}\xspace}
\newcommand {\mean}[1]   {\ensuremath{\left\langle #1 \kern-0.1em\right\rangle}\xspace}

\newcommand {\Raa}       {\ensuremath{R_\mathrm{AA}}\xspace}

\newcommand {\pp}        {\mbox{$\mathrm {p\kern-0.05em p}$}\xspace}
\newcommand {\ppBoldMath} {\mbox{$\mathrm { \mathbf p\kern-0.05em \mathbf p }$}\xspace}

\newcommand {\PbPb}      {\ensuremath{\mbox{Pb--Pb}}\xspace}

\newcommand {\pPb}       {\ensuremath{\mbox{p--Pb}}\xspace}

\newcommand {\MeanNpart} {\mbox{\ensuremath{< \kern-0.15em N_\mathrm{part} \kern-0.15em >}}}

\newcommand{\RpPb}{\ensuremath{R_\mathrm{pPb}}\xspace}
\newcommand{\RpA}{\ensuremath{R_\mathrm{pA}}\xspace}

\newcommand {\mass}     {\mbox{\rm MeV$\kern-0.15em /\kern-0.12em c^2$}}

\newcommand {\mmom}     {\mbox{\rm MeV$\kern-0.15em /\kern-0.12em c$}\xspace}
\newcommand {\gmom}     {\mbox{\rm GeV$\kern-0.15em /\kern-0.12em c$}\xspace}
\newcommand {\mmass}    {\mbox{\rm MeV$\kern-0.15em /\kern-0.12em c^2$}\xspace}
\newcommand {\gmass}    {\mbox{\rm GeV$\kern-0.15em /\kern-0.12em c^2$}\xspace}

\newcommand {\dg}       {\mbox{$\kern+0.1em ^\circ$}\xspace}

\newcommand{\Dz}{\Dzero}

\newcommand{\Lc}        {\mbox{$\mathrm {\Lambda_{c}^{+}}$}\xspace}

\newcommand{\rmLambdas}         {\mbox{$\mathrm {\Lambda \kern-0.2em + \kern-0.2em \overline{\Lambda}}$}\xspace}

\newcommand{\Kzs}               {\ensuremath{\mathrm {K^0_S}}\xspace}

\newcommand{\Jpsi}	{\ensuremath{\mathrm{J/\psi}}\xspace}
\newcommand{\LKs} {\ensuremath{\Lambda/\Kzs}\xspace}

\newcommand{\LcD}   {\ensuremath{\Lc/\Dz}\xspace}

\newcommand{\GeV}{\ensuremath{\mathrm{GeV}}\xspace}
\newcommand{\TeV}{\ensuremath{\mathrm{TeV}}\xspace}
\newcommand{\gevc}{\ensuremath{\mathrm{GeV}/c}\xspace}
\newcommand{\GeVc}{\gevc}

\newcommand{\Dzero}{\ensuremath{\mathrm{D}^0}\xspace}

\newcommand{\Dsubs}{\ensuremath{\rm D^+_\mathrm{s}}\xspace}
\newcommand{\Bsubs}{\ensuremath{\rm B^0_\mathrm{s}}\xspace}

\newcommand{\Bplus}{\ensuremath{\rm B^+}\xspace}

\newcommand{\sqrtsseven}{\ensuremath{\sqrt{s}=7\,\TeV}\xspace}

\newcommand{\sqrtsfive}{\ensuremath{\sqrt{s}=5.02\,\TeV}\xspace}

\newcommand{\sqrtsNNfive}{\ensuremath{\sqrt{s_\mathrm{NN}}=5.02\,\TeV}\xspace}
\newcommand{\sqrtsNNtwo}{\ensuremath{\sqrt{s_\mathrm{NN}}=2.76\,\TeV}\xspace}

\newcommand{\figref}[1]{Fig.~\ref{#1}}

\newcommand{\Figref}[1]{Figure~\ref{#1}}

\newcommand{\eqref}[1]{Eq.~(\ref{#1})}

\begin{document}
\title{Open heavy-flavour production in heavy-ion collisions at the LHC}

\author{Jeremy Wilkinson on behalf of the ALICE, ATLAS, CMS and LHCb Collaborations}

\address{INFN Sezione di Bologna, Italy}

\ead{jeremy.wilkinson@cern.ch}
\begin{abstract}
This work presents the latest results on open heavy-flavour production in proton--proton, proton--lead, lead--lead and xenon--xenon collisions from the ALICE, ATLAS, CMS and LHCb Collaborations at the LHC.
\end{abstract}

\section{Physics Motivation}
The production of heavy-flavour hadrons (those containing charm or beauty quarks) is one of the key probes of the colour-deconfined and strongly interacting state of matter, the Quark--Gluon Plasma (QGP) that is formed in ultrarelativistic heavy-ion collisions~\cite{andronic}. Due to their large masses with respect to the QCD energy scale ($m_\mathrm{c,b} \gg \Lambda_\mathrm{QCD}$), they are formed at early times in hard (high-$Q^2$) partonic scatterings. This means that they experience the full evolution of the system before decaying, and so studying their behaviour in heavy-ion collisions gives a unique handle on the transport coefficients of the medium.

In the context of heavy-ion physics, it is necessary to study multiple different collision systems to fully understand the observed effects. Proton--proton collisions are used as a baseline for cross-section measurements in the absence of a medium, and can also be used to test predictions made by perturbative QCD (pQCD) calculation frameworks. Minimum-bias proton--lead collisions can also be compared with lead--lead collisions in order to disentangle the role of cold nuclear matter (CNM) effects that arise simply from the presence of a nucleus in the initial state, such as nuclear shadowing, from those occurring due to the presence of a deconfined medium. Lead--lead and, more recently, xenon--xenon collisions at the LHC, form the basis of the QGP studies among the LHC experiments, and comparisons between the two systems at similar numbers of binary nucleon--nucleon collisions ($N_\mathrm{coll}$) can provide an insight into the path length dependence of energy loss in the medium.

Charmed baryon production has also recently become an important topic in heavy-ion physics. Previous studies in the strangeness sector  have shown an enhancement in the baryon-to-meson ratio of $\Lambda$ baryons compared with kaons due to the effect of recombination of strange quarks with light quarks in the medium. Similar studies  in the charm and beauty sectors will shed light on the fragmentation and hadronisation mechanisms of heavy quarks.

Heavy flavours are measured in the four LHC experiments using a variety of techniques: the reconstruction of electrons and muons from the semileptonic decays of open heavy-flavour hadrons (ALICE, ATLAS), where it is possible to separate the beauty contribution from that of charm; direct reconstruction of charmed mesons and baryons (ALICE, ATLAS, CMS, and LHCb), which retains the full kinematic information of the original particle; and the reconstruction of non-prompt \Jpsi mesons which originate from the decay of beauty particles.

The main observables used in the context of heavy-ion collisions are the nuclear modification factor, $R_\mathrm{AA}$, and the Fourier coefficients $v_\mathrm{n}$ of the azimuthal distribution of particle momenta. The $R_\mathrm{AA}$ is the ratio of the production rate in nuclear collisions to that in proton--proton collisions at the same $\sqrt{s}$, scaled by the number of binary nucleon--nucleon collisions. A similar quantity, $R_\mathrm{pA}$, is defined for proton--nucleus collisions, and is the ratio of the p--A and pp cross sections scaled by the mass number of the nucleus. If the \RpA is equal to unity then this implies that there are no CNM effects present; an \Raa of less than 1 is evidence of suppression, which at intermediate \pT is typically a sign of partonic energy loss in the medium. The energy loss in the medium is expected to be dependent on the mass and colour charge of particles due to the so-called `dead cone effect', leading to an expected hierarchy of $\Delta E(\mathrm{g}) > \Delta E(\mathrm{u,d,s}) > \Delta E(\mathrm{c}) > \Delta E(\mathrm{b})$, and so $R_\mathrm{AA}(\mathrm{g}) < R_\mathrm{AA}(\mathrm{u,d,s}) < R_\mathrm{AA}(\mathrm{c}) < R_\mathrm{AA}(\mathrm{b})$.

The $v_\mathrm{n}$ coefficients can be expressed from a Fourier decomposition of the angular distribution of particles, with each \vn representing the coefficient for the $n^{\mathrm{th}}$ harmonic. The main notable \vn parameters are $v_1$ (``direct flow''), which gives insight into magnetic field behaviour in the medium; $v_2$ (``elliptic flow''), which shows the degree to which a spatial asymmetry and resulting pressure gradient in the initial state translates to a momentum asymmetry in the final state, thus revealing the degree of collective behaviour in the expansion; and $v_3$ (``triangular flow''), which arises mainly due to fluctuations in the nuclear overlap and provides further information on the collective behaviour of particles in the medium.

\section{Measurements in small collision systems}

\begin{figure}[h!]
\centering
\resizebox{.99\textwidth}{!}{\includegraphics[height=3cm]{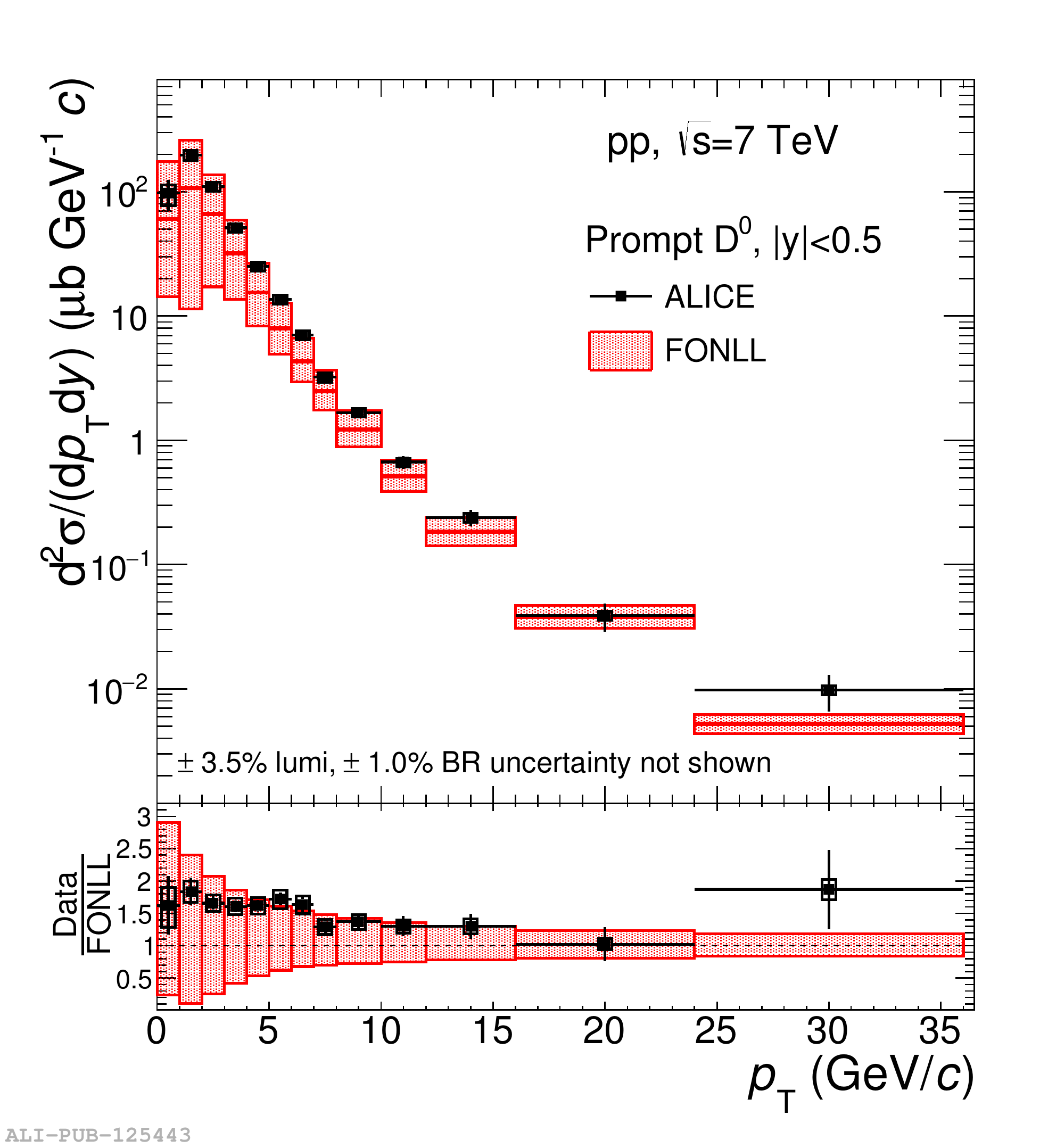}
\includegraphics[height=3cm]{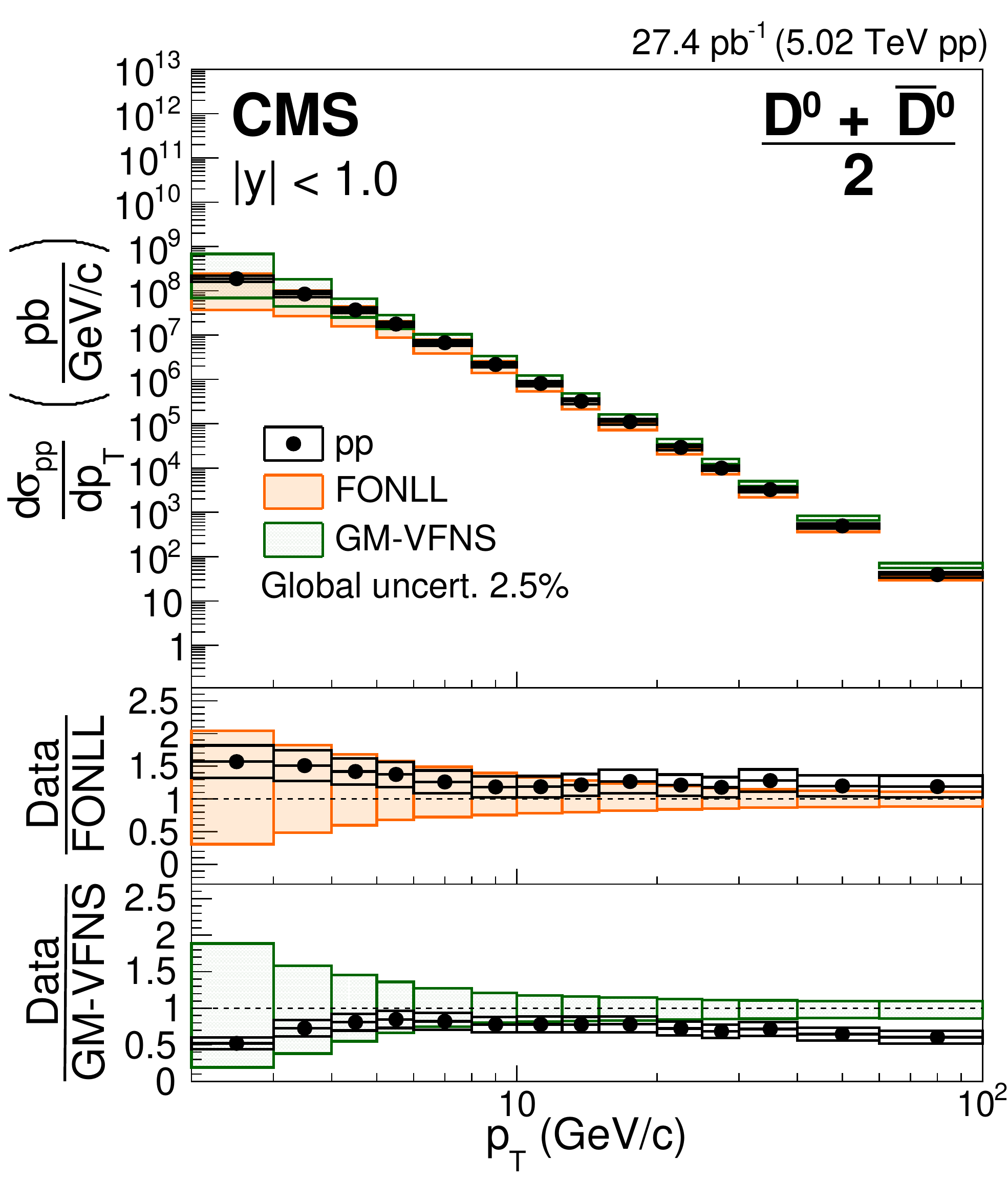}
\includegraphics[height=3cm]{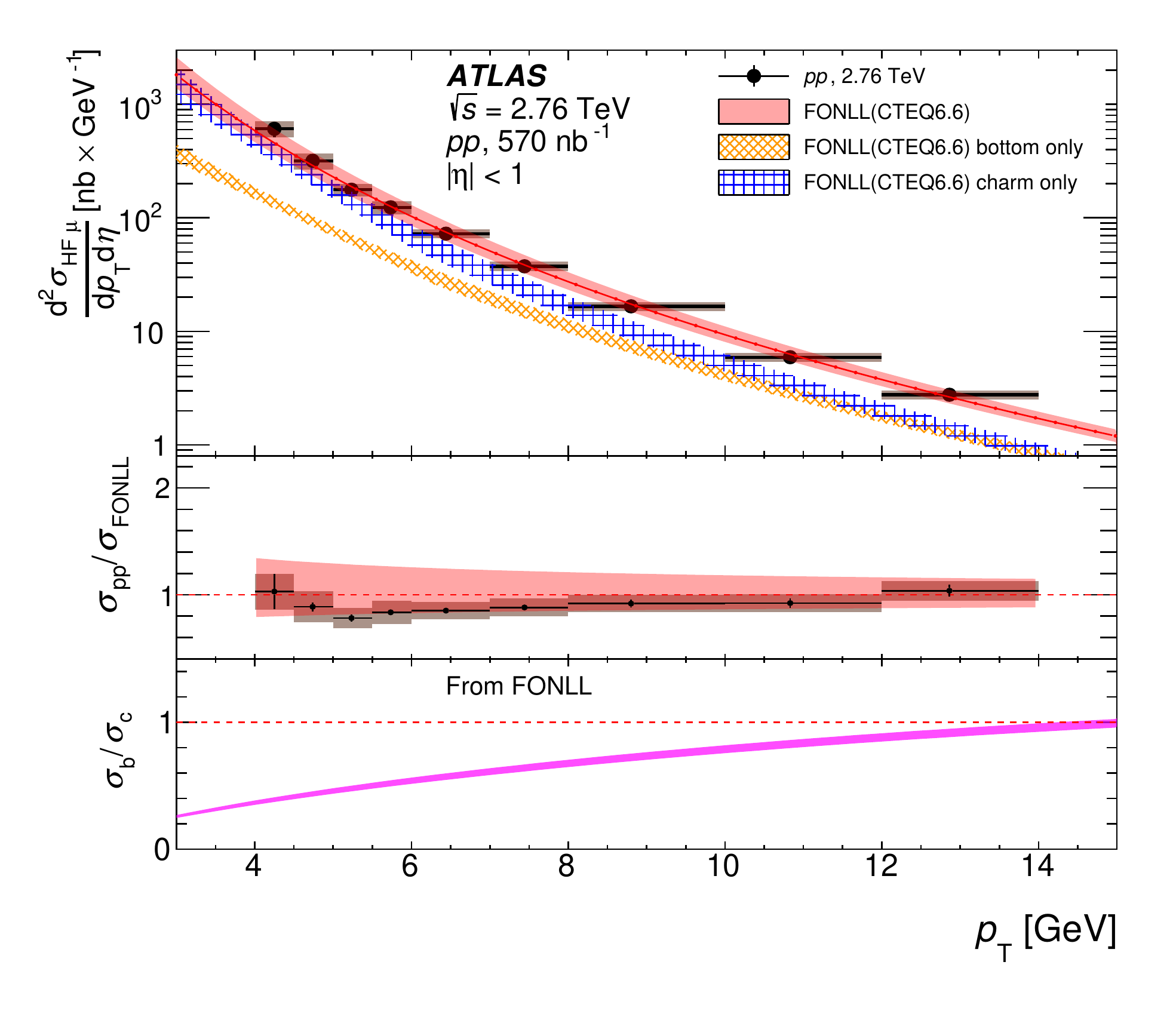}
}
\caption{Left: ALICE measurement of the \Dz-meson production cross section in pp collisions at \sqrtsseven, compared with FONLL pQCD calculations~\cite{alicepp}. Middle: CMS measurement of the \Dz-meson production cross section in pp collisions at \sqrtsfive, compared with FONLL and GM-VFNS calculations~\cite{cmsraa}. Right: ATLAS measurement of the inclusive heavy-flavour decay muon production cross section~\cite{atlaspbpb}.}
\label{fig:pp}
\end{figure}

Proton--proton collisions are the main baseline for heavy-ion measurements. \Figref{fig:pp} shows the cross section measurements for \Dz mesons in pp collisions from ALICE~\cite{alicepp} and CMS~\cite{cmsraa}, and for inclusive heavy-flavour decay muons from ATLAS~\cite{atlaspbpb}. In each case the measurements are described within uncertainties by perturbative QCD models such as FONLL; however, the D mesons measured by CMS and ALICE tend to populate the upper region of the theoretical uncertainty band.

\begin{figure}[h!]
\centering
\resizebox{.8\textwidth}{!}{%
\includegraphics[height=3cm]{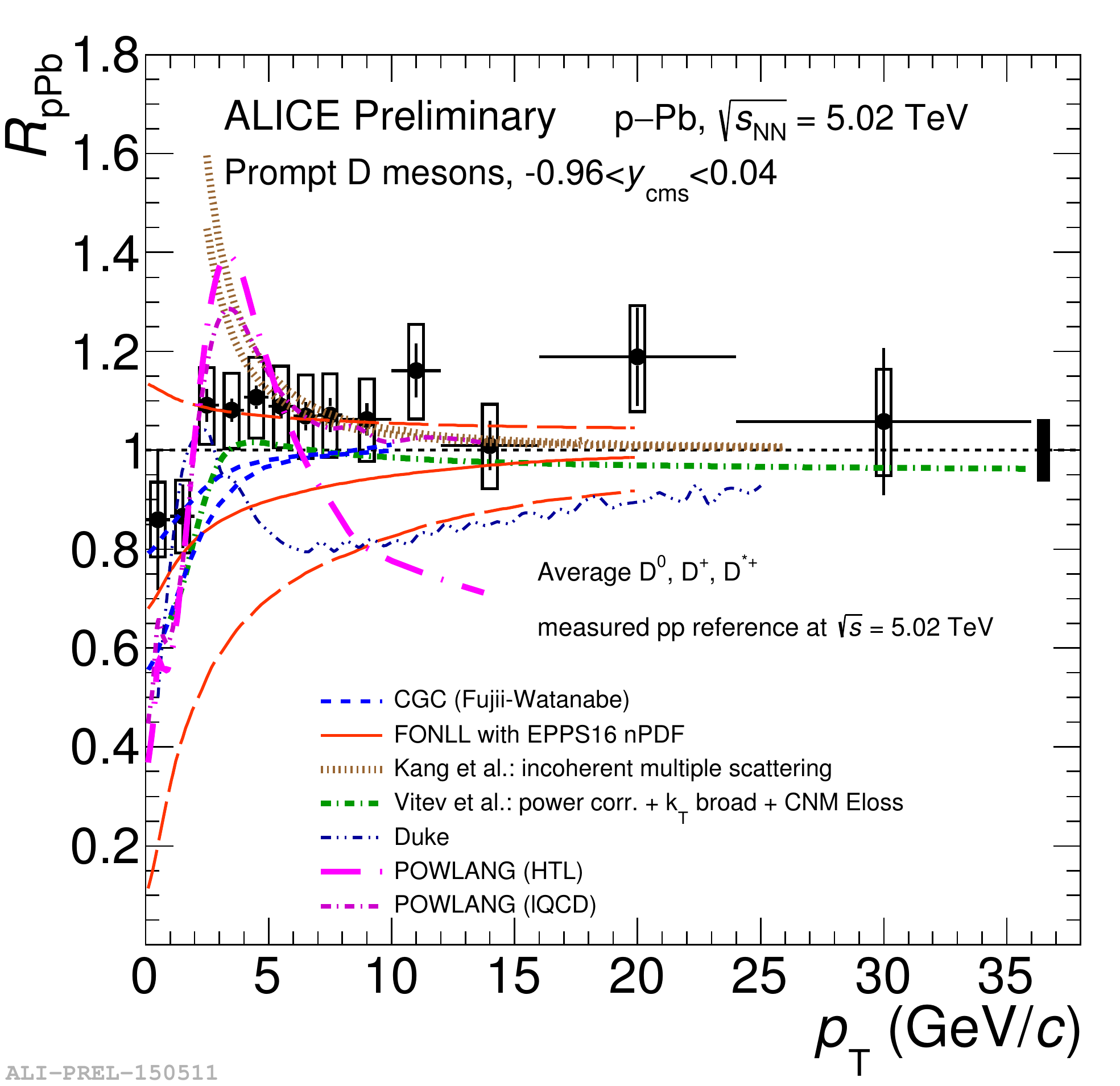}
\includegraphics[height=3cm]{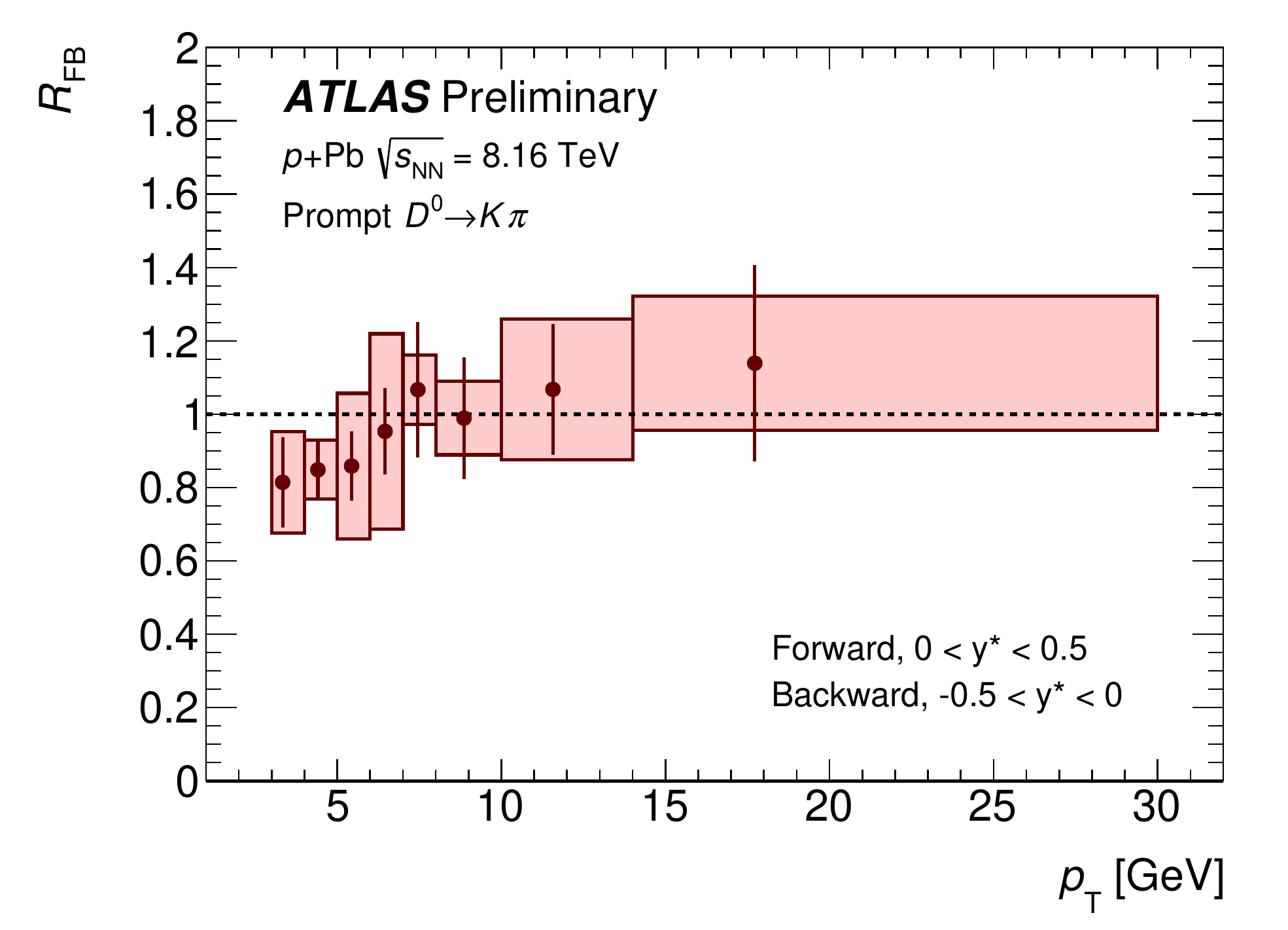}
}
\caption{Left: ALICE preliminary measurement of \RpPb of D mesons at \sqrtsNNfive using Run-2 data. Right: ATLAS preliminary meaurement of forward-to-backward production ratio of \Dzero mesons in \pPb collisions at mid-rapidity.}
\label{fig:pa}
\end{figure}

\Figref{fig:pa} illustrates measurements from ALICE and ATLAS for D-meson production in proton--nucleus collisions. The ALICE measurement shows the \RpPb of non-strange D mesons at mid-rapidity, which is compatible with both unity and models within uncertainties, thus implying that any modification seen in nucleus--nucleus collisions is not significantly due to CNM effects. The plot from ATLAS shows the $R_\mathrm{FB}$, or ratio of forward to backward production, at mid-rapidity for \pPb collisions. Here, the ratio is equal to unity, implying that there is no significant rapidity dependence of charm-quark production in this kinematic region. The LHCb Collaboration has also measured the \Dzero-meson production cross section in its fixed target p--Ar campaign, finding that the p--Ar cross section is similar to that in pp collisions at low \pT. The results in both collision systems are described well by theoretical model calculations.

\begin{figure}[h!]
\centering
\resizebox{.8\textwidth}{!}{%
\includegraphics[height=3cm]{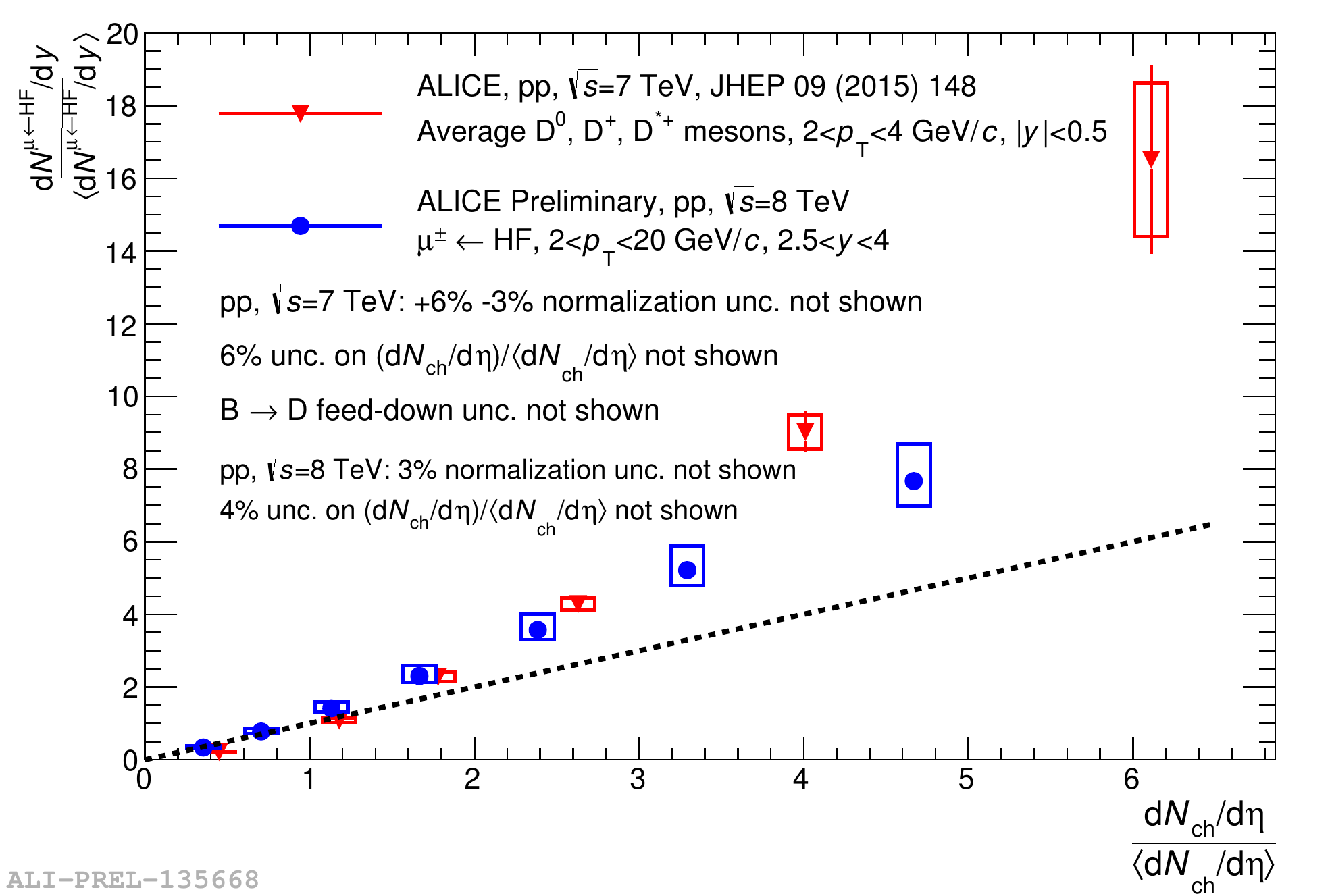}
\includegraphics[height=3cm]{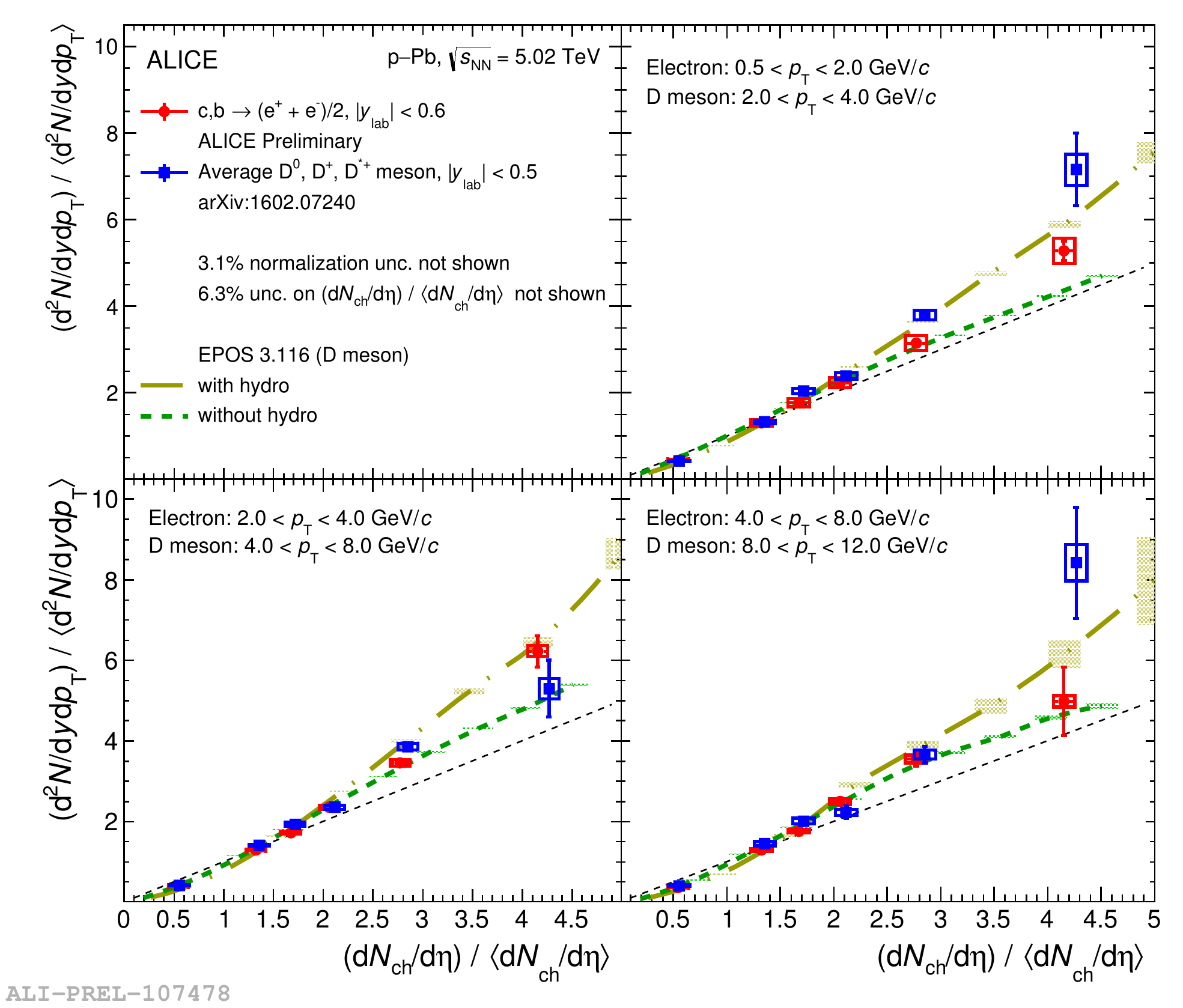}
}
\caption{Multiplicity dependence of heavy-flavour particle production in pp and \pPb collisions measured by the ALICE Collaboration. Left: D-meson and inclusive heavy-flavour decay muon production in proton--proton collisions. Right: D-meson and inclusive heavy-flavour decay electron production in \pPb collisions.}
\label{fig:multdep}
\end{figure}

Further measurements are made in small systems to determine the multiplicity dependence of particle production, allowing the role of multi-parton interactions (MPI) to be studied. \Figref{fig:multdep} depicts the ALICE measurements for the multiplicity dependence of D mesons and heavy-flavour decay muons in pp collisions and for D mesons and heavy-flavour decay electrons in p--Pb collisions. In each case, the `self-normalised yield' is plotted, where the yield per event in a given multiplicity class is divided by the yield per event averaged over all multiplicities. The $x$-axis shows the equivalent value for overall charged particle production; a linear rise (indicated by the black dashed $y=x$ lines) would therefore indicate a simple correlation between hard and soft particle production processes. In fact, a slightly faster-than-linear trend is seen at higher multiplicities in both collision systems and at all rapidities, implying that the presence of multiple hard interactions is more relevant at high multiplicity. The results are better described by the EPOS model in p--Pb collisions when viscous hydrodynamic effects are considered in the calculations than when they are not, potentially implying that some degree of collectivity is present here.

\begin{figure}[h!]
\centering
\resizebox{.98\textwidth}{!}{%
\includegraphics[height=3cm]{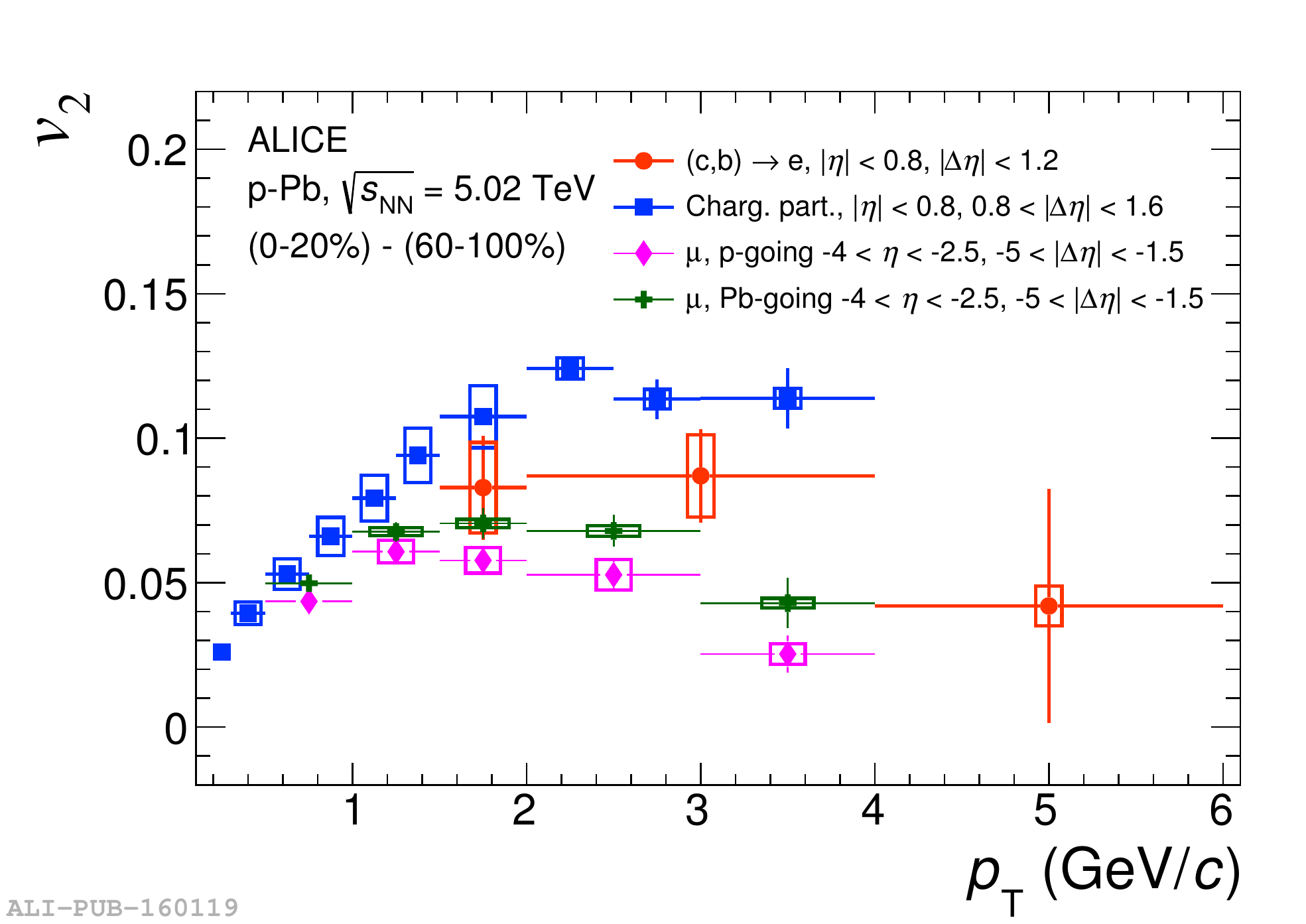}
\includegraphics[height=3cm]{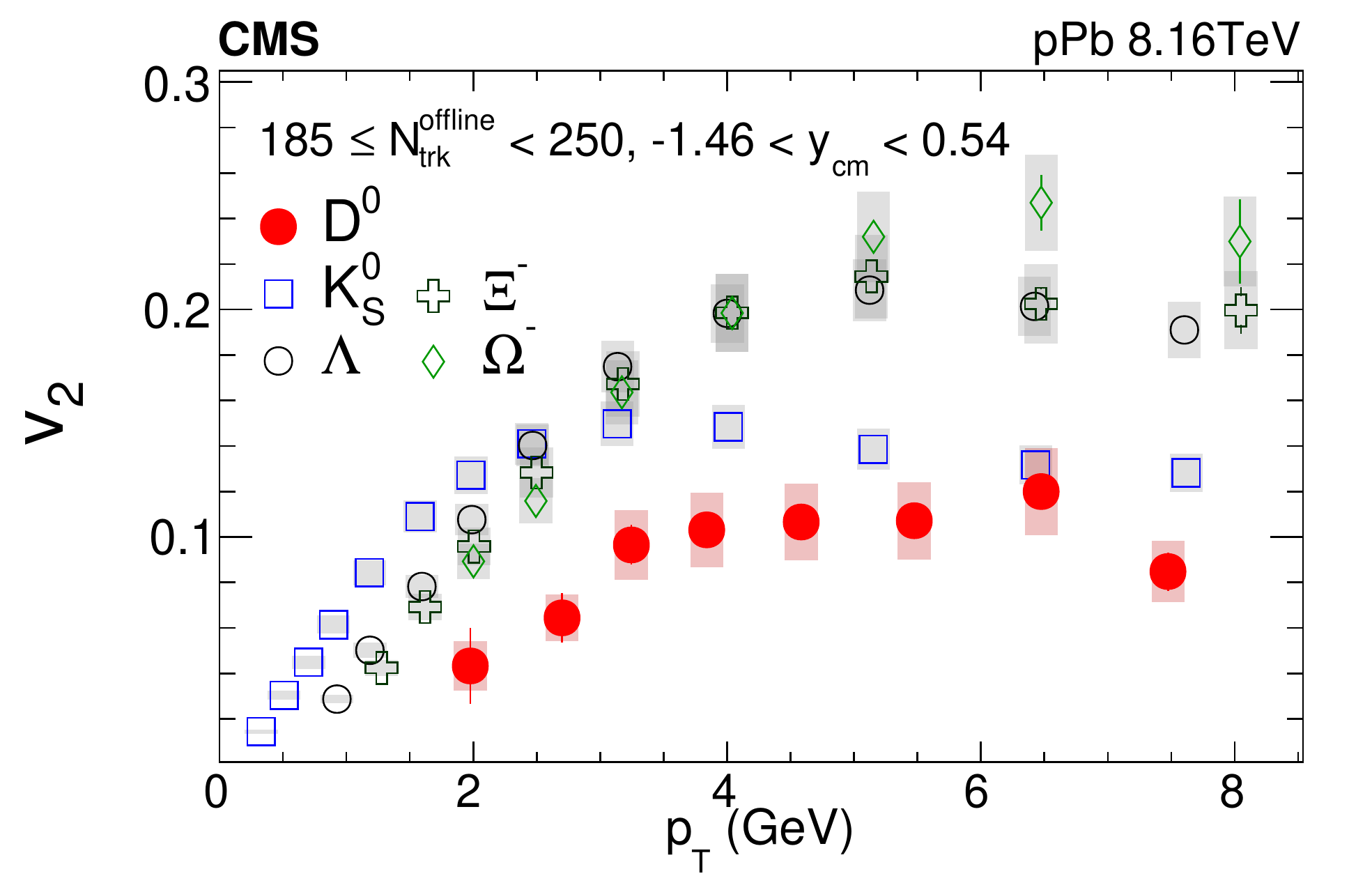}
\includegraphics[height=3cm]{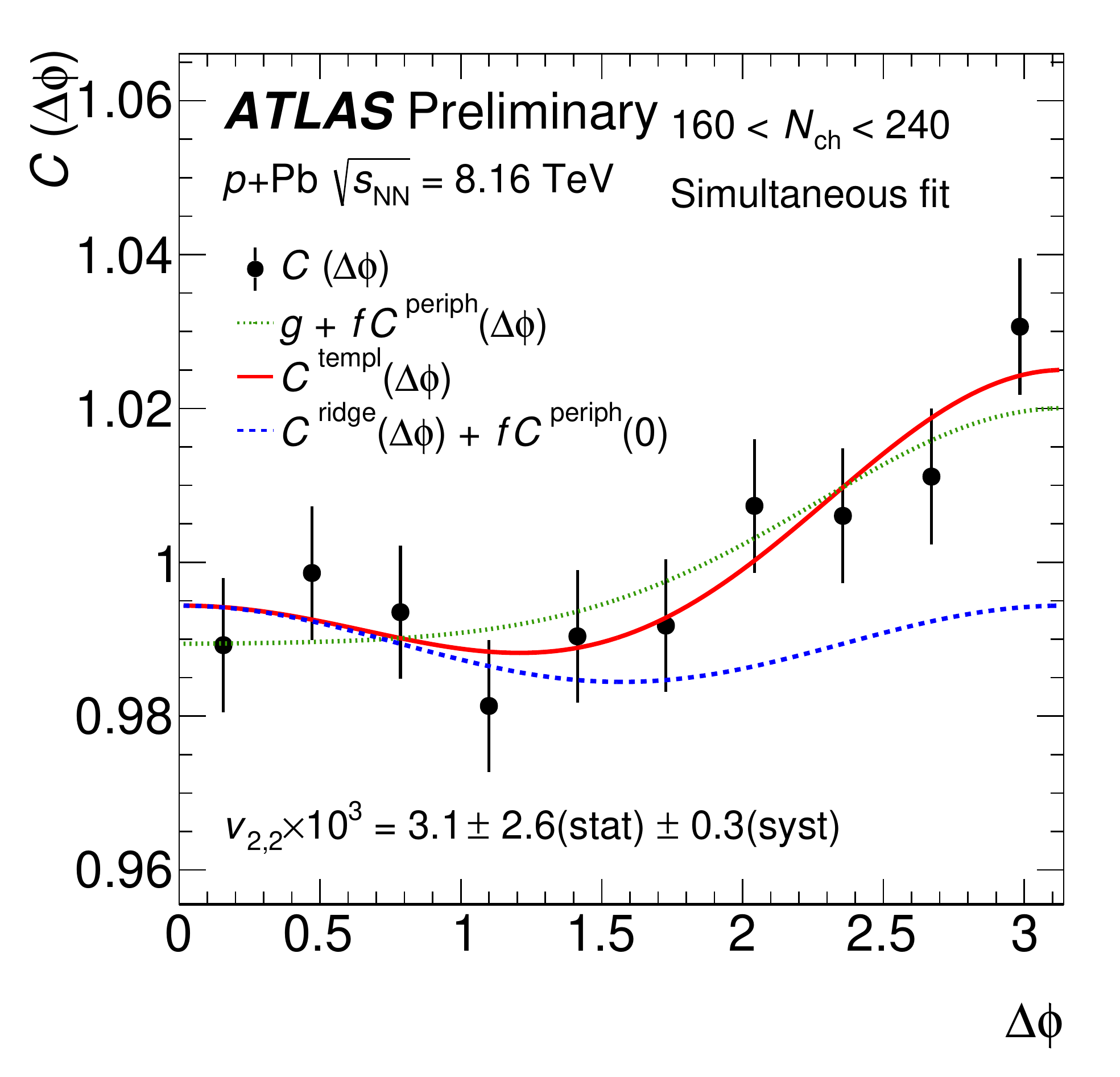}
}
\caption{Left: ALICE measurement of the $v_2$ coefficient of inclusive heavy-flavour decay electrons and muons, compared with charged particles~\cite{alicev2pa}. Middle: CMS measurement of \Dz-meson $v_2$ coefficient compared with strange mesons and baryons~\cite{cmsv2pa}. Right: ATLAS measurement of two-particle correlation function for D mesons, showing the extracted $v_{2,2}$ coefficient.}
\label{fig:pav2}
\end{figure}

This can be further demonstrated by considering the elliptic flow, which is studied through the means of long-ranged angular correlations in p--Pb collisions. The results for CMS, ALICE and ATLAS are shown in~\figref{fig:pav2}, where a significant positive flow parameter is seen by all three experiments. The CMS measurement shows a smaller $v_2$ parameter for \Dz mesons than for strange mesons and baryons; a similar effect is seen when comparing the ALICE results for heavy-flavour decay leptons with overall charged particle production. The heavy-flavour decay muons and electrons from ALICE are measured in three separate rapidity classes; no significant dependence on rapidity is seen for the $v_2$ coefficient in these kinematical ranges.

\section{Measurements in Pb--Pb and Xe--Xe collisions}

\begin{figure}[h!]
\centering
\resizebox{.95\textwidth}{!}{%
\includegraphics[height=3cm,trim=0 0 720 0,clip]{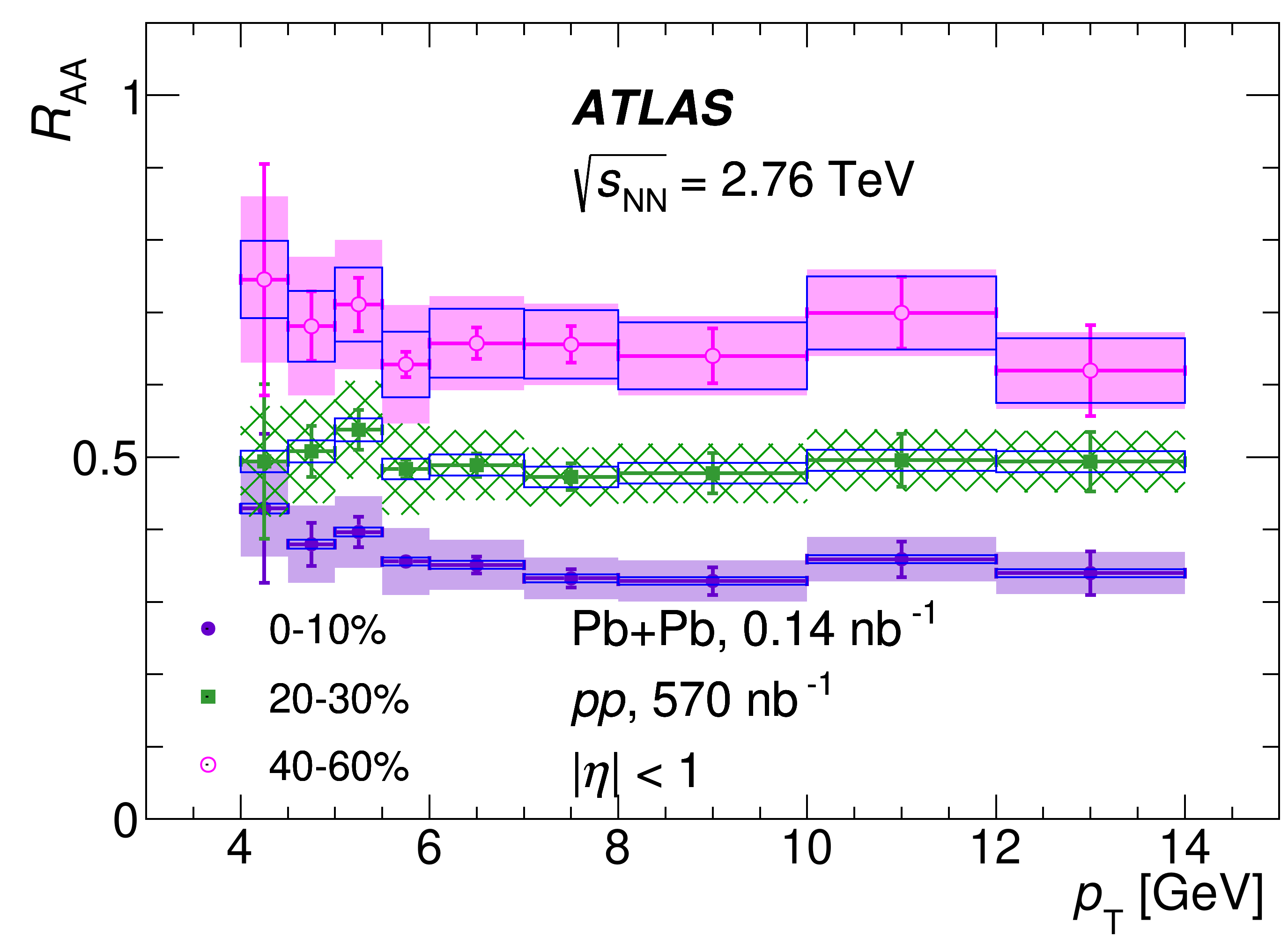}
\includegraphics[height=3cm]{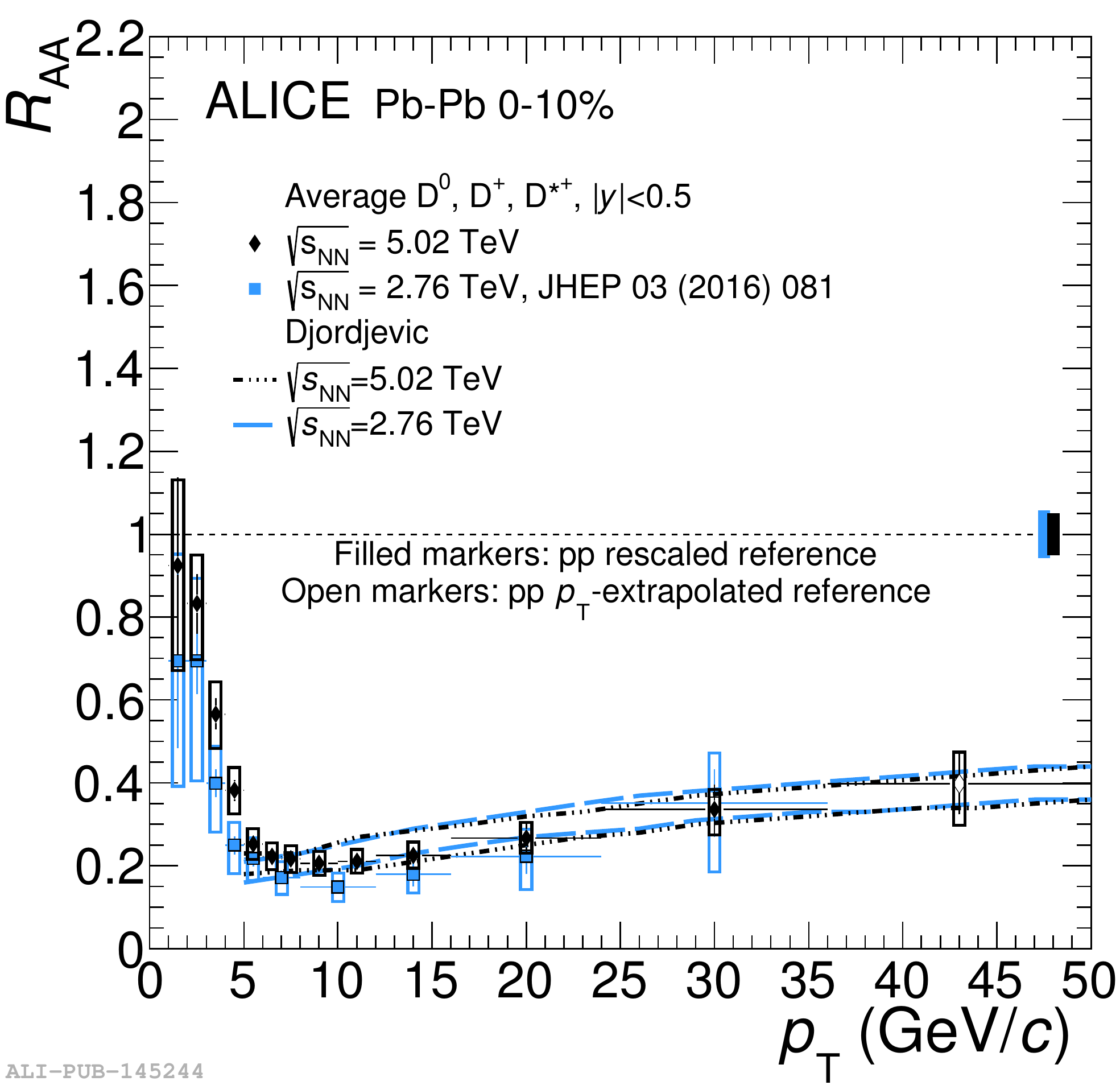}
\includegraphics[height=3cm]{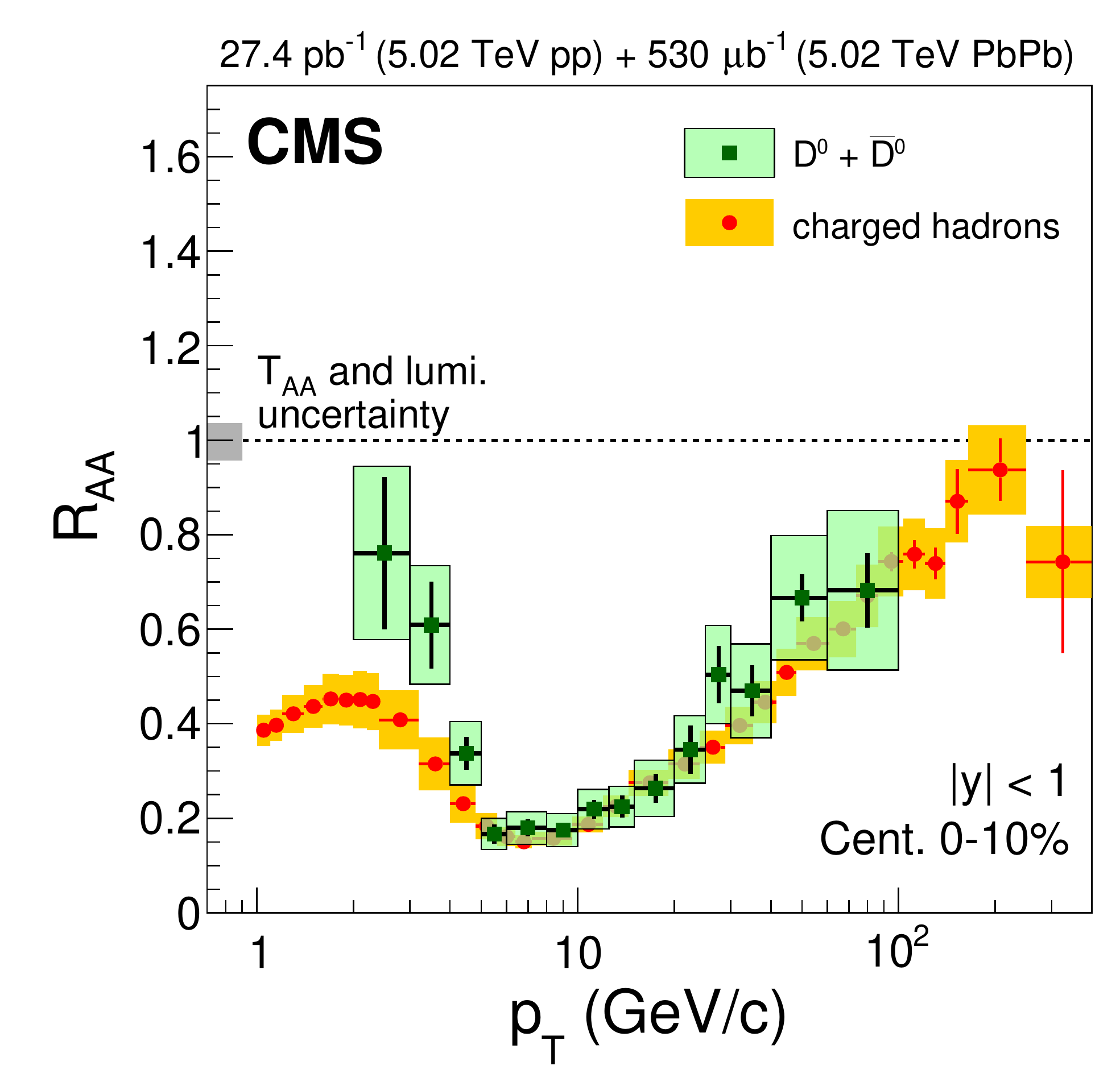}
}
\caption{Measurements of the \Raa of non-strange heavy hadrons in \PbPb collisions at the LHC. Left: ATLAS measurement of inclusive heavy-flavour decay muons in three centrality classes~\cite{atlaspbpb}. Middle: ALICE measurement of non-strange D-meson \Raa at \sqrtsNNtwo and \sqrtsNNfive~\cite{aliceraa}. Right: CMS measurement of \Dz mesons compared with charged hadrons~\cite{cmsraa}.}
\label{fig:nonstrangeheavy}
\end{figure}

The \Raa of non-strange heavy-flavour hadrons is shown in~\figref{fig:nonstrangeheavy}. The left plot shows the ATLAS measurement of inclusive heavy-flavour decay muons in three centrality classes~\cite{atlaspbpb}. Here, it can be seen that there is a significant suppression of heavy-flavour particle production at intermediate \pT, and that the degree of suppression increases (\Raa decreases) when considering more central collisions. The middle plot shows the ALICE measurement of non-strange D mesons at \sqrtsNNtwo and \sqrtsNNfive for 0--10\% central collisions, compared with energy loss models~\cite{aliceraa}. A hint of a difference in the \Raa based on the collision energy can be seen at low \pT; however, this is not significant with the current measurement uncertainties. The \pT distribution at high \pT is well described by the models. The right plot shows the equivalent measurement by CMS for \Dz mesons, compared with the \Raa of charged hadrons, which are dominated by light-flavour particles~\cite{cmsraa}. While there is no difference seen between D mesons and lighter particles at intermediate to high \pT, a significantly lesser degree of suppression is seen for D mesons at $\pT < 5\,\GeVc$.

\begin{figure}[h!]
\centering
\resizebox{.8\textwidth}{!}{%
\includegraphics[height=3cm]{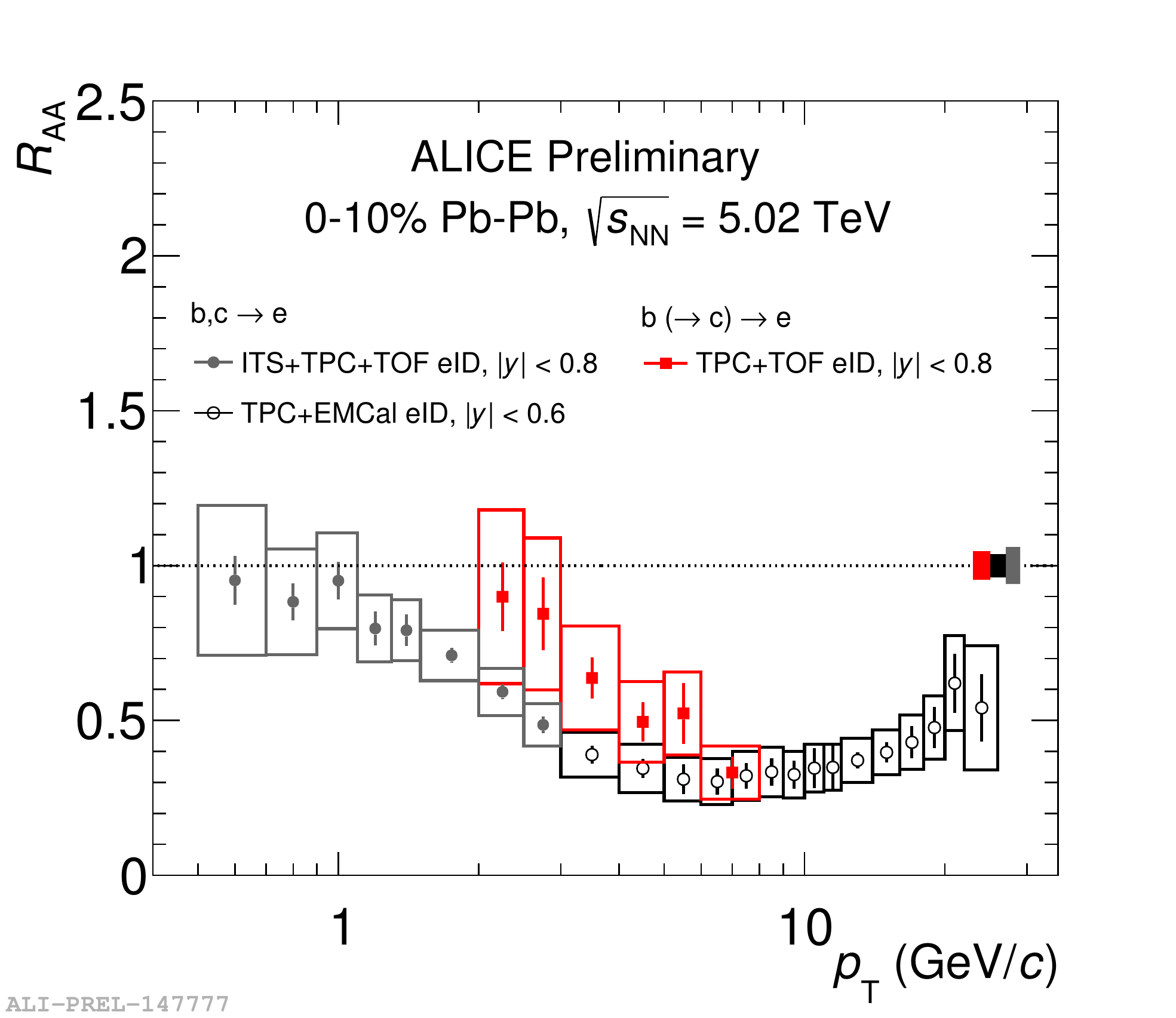}
\includegraphics[height=3cm]{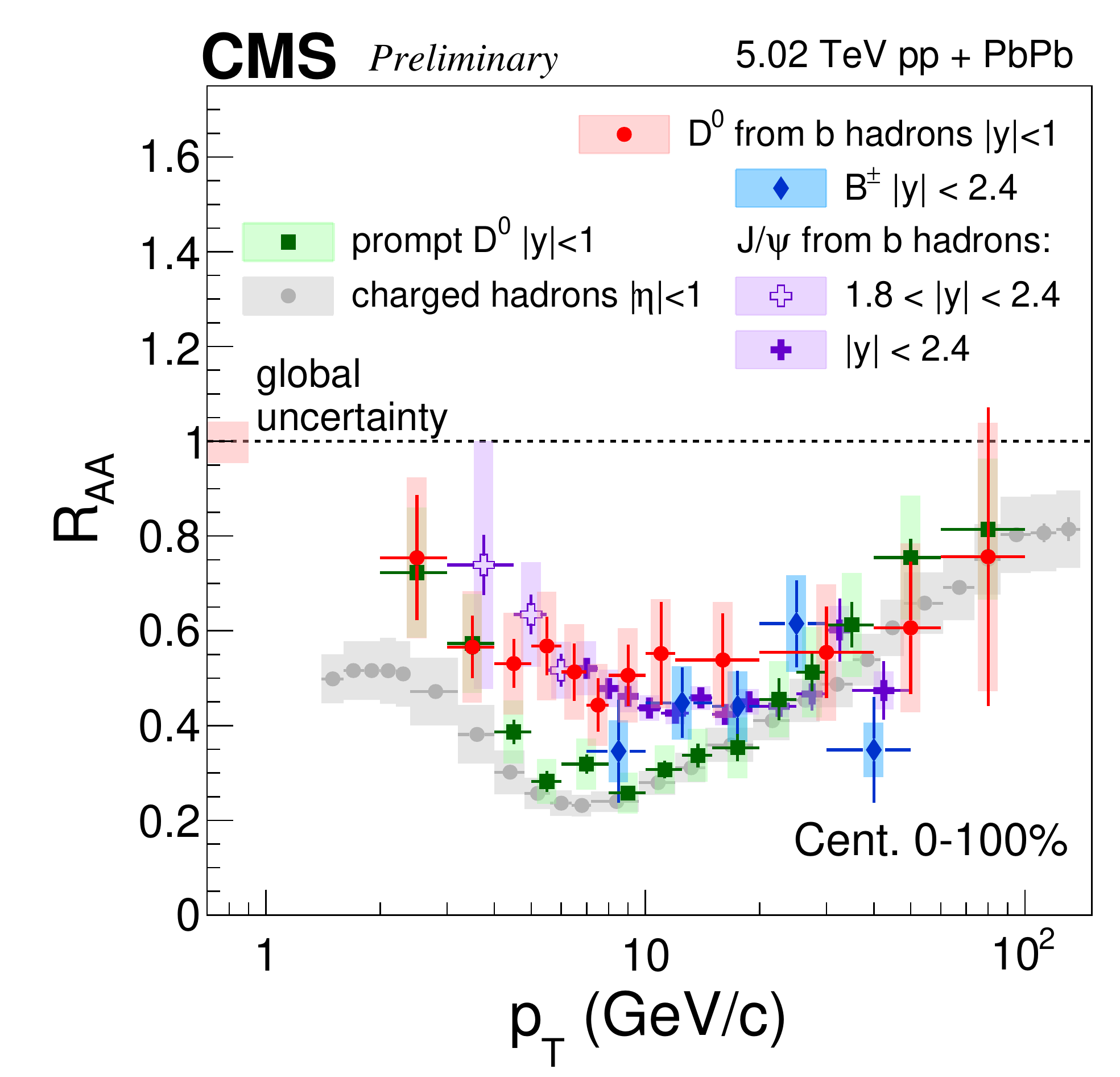}
}
\caption{Measurements of the \Raa of beauty hadrons. Left: ALICE measurement of beauty-decay electrons compared with inclusive heavy-flavour decay electrons. Right: CMS measurement of non-prompt \Jpsi and of \Bplus mesons compared with \Dz mesons and charged hadrons.}
\label{fig:beauty}
\end{figure}

The beauty \Raa sheds further light on the mass dependence of partonic energy loss. This is shown in~\figref{fig:beauty}: on the left, for electrons from beauty-hadron decays in ALICE, compared with inclusive heavy-flavour decay electrons; and on the right for both non-prompt \Jpsi (which originate from B decays) and \Bplus mesons in CMS. In each case, it can be seen that the contribution to the inclusive spectra by beauty hadrons is still suppressed, but to a lesser degree than for charmed hadrons, partially confirming the prediction that beauty hadrons suffer less energy loss than charmed hadrons.

\begin{figure}[h!]
\centering
\resizebox{.7\textwidth}{!}{%
\includegraphics[height=3cm]{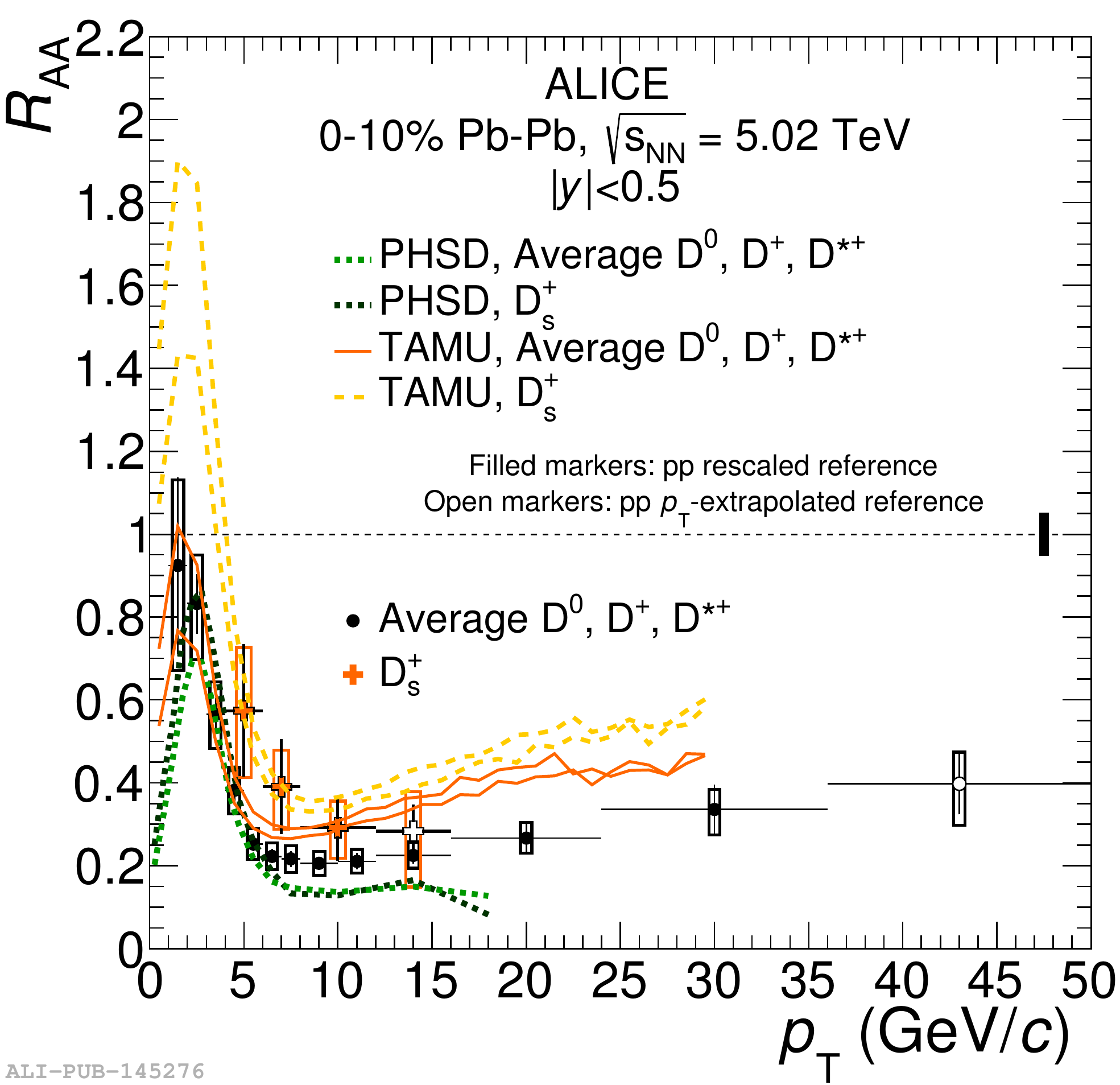}
\includegraphics[height=3cm]{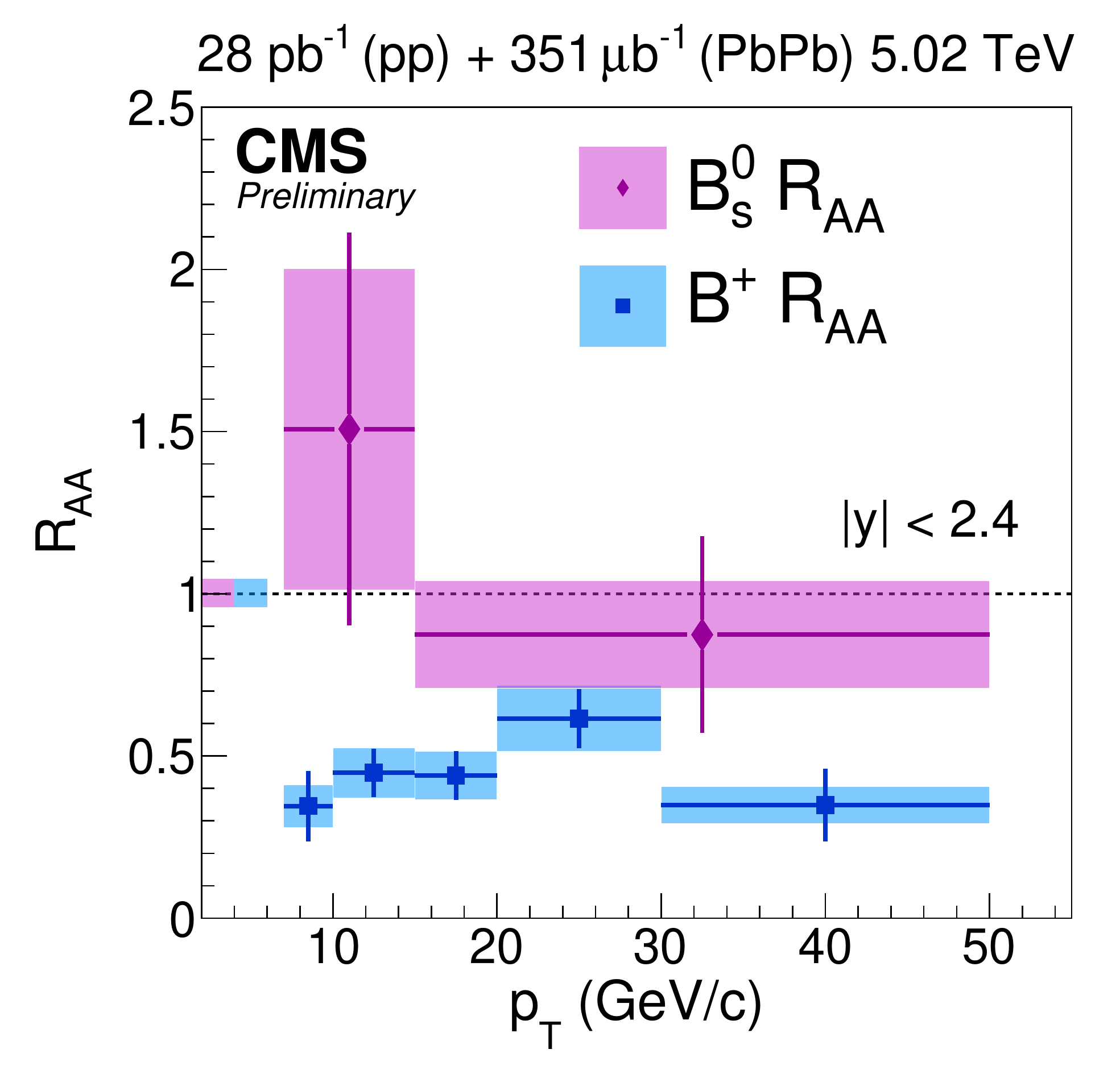}

}
\caption{Measurements of strange heavy-meson production in \PbPb collisions at \sqrtsNNfive. Left: ALICE measurement of the \Raa of \Dsubs-meson production compared with non-strange D mesons and models~\cite{aliceraa}. Right: CMS preliminary measurement of the \Raa of \Bsubs mesons compared with \Bplus mesons.}
\label{fig:strangeheavy}
\end{figure}

It is also instructive to consider the production of strange heavy mesons in heavy-ion collisions. The enhancement of strangeness production is seen as a potential signature of QGP formation, and therefore a modification of \Dsubs-meson and \Bsubs-meson production with respect to non-strange D and B mesons could be a sign of recombination of charm quarks with strange quarks in the medium. The ALICE measurement of the \Raa of non-strange and strange D mesons~\cite{aliceraa}, and the CMS measurement of \Bsubs and \Bplus mesons, are shown in~\figref{fig:strangeheavy}. In both cases, the heavy mesons containing a strange quark appear to be less suppressed than those without, although the measurement uncertainties are still large enough that it is difficult to draw a firm conclusion. The TAMU model~\cite{tamu}, which considers only elastic energy loss processes, describes the \Dsubs \Raa better than the PHSD model~\cite{phsd}, which also includes gluon radiation.
	
\begin{figure}[h!]
\centering
\resizebox{.9\textwidth}{!}{%
\includegraphics[height=2.7cm]{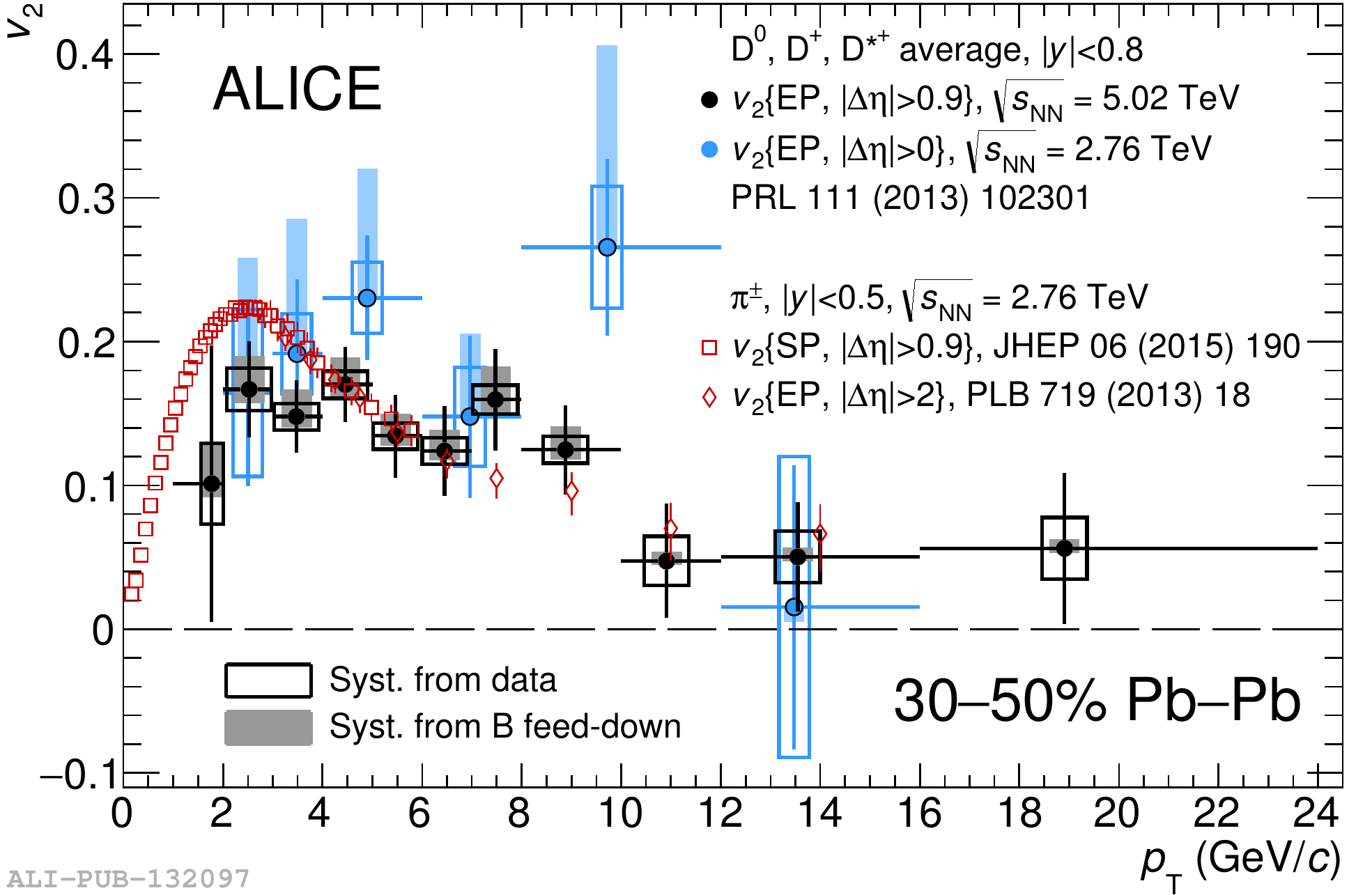}
\includegraphics[height=3cm]{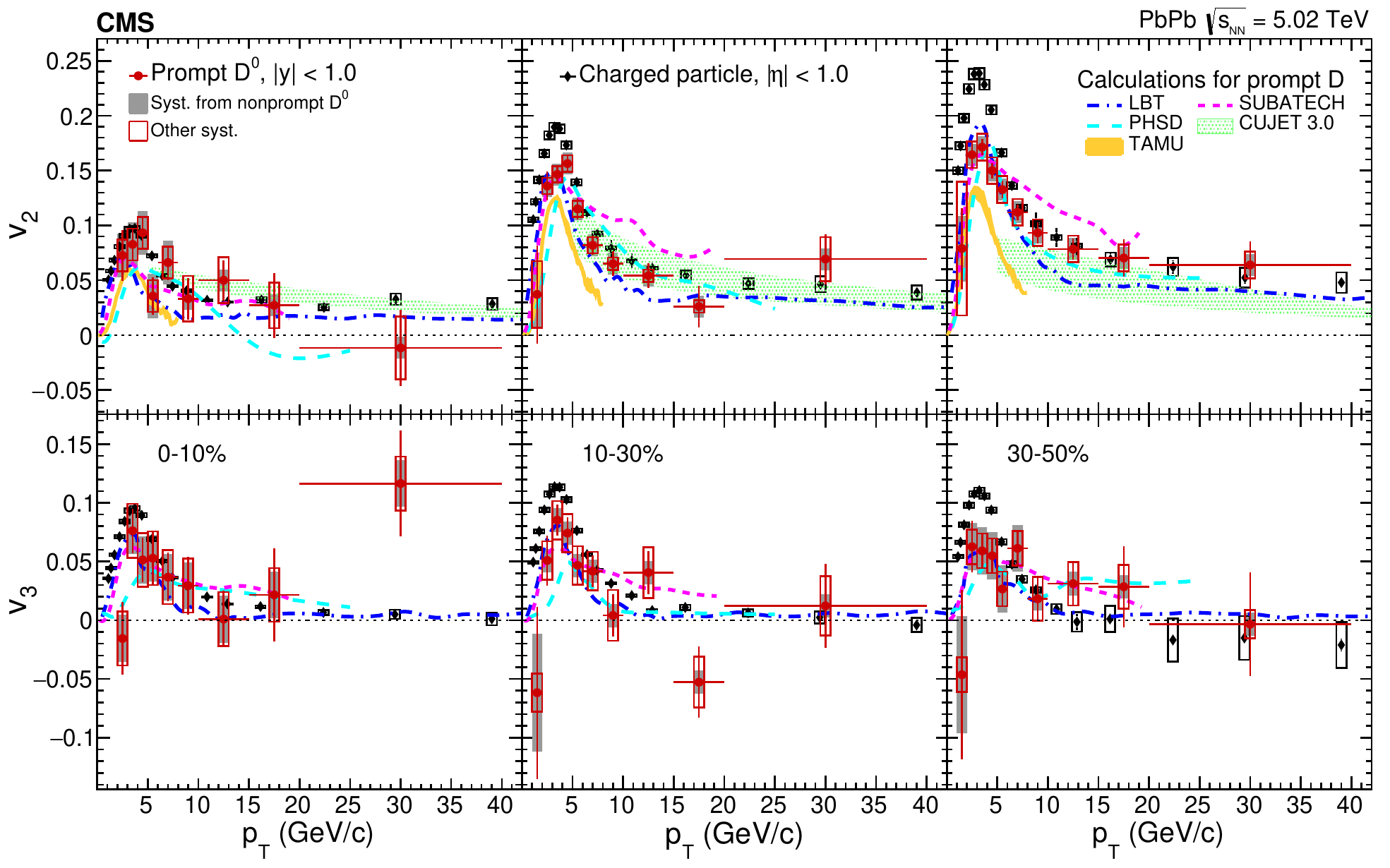}
}
\caption{Left: elliptic flow of D mesons measured by the ALICE Collaboration at \sqrtsNNtwo and \sqrtsNNfive, compared with charged particles~\cite{alicev2}. Right: elliptic ($v_2$) and triangular ($v_3$) flow of \Dzero mesons measured by the CMS Collaboration compared with charged particles and models in three centrality intervals~\cite{cmsv2}.}
\label{fig:pbpbflow}
\end{figure}

Measurements of the flow parameters of charmed hadrons in \PbPb collisions is shown in~\figref{fig:pbpbflow}. The left plot displays the elliptic flow ($v_2$) measurements for D mesons and charged pions by the ALICE Collaboration in 30--50\% central Pb--Pb collisions at \sqrtsNNfive and \sqrtsNNtwo~\cite{alicev2}; the right plot shows the elliptic and triangular ($v_3$) flow parameters for \Dz mesons measured by the CMS Collaboration at \sqrtsNNfive~\cite{cmsv2}. The $v_2$ of D mesons is largely consistent with that of charged pions, suggesting that they participate in the collective expansion in the medium to a similar degree, and there appears to be no significant dependence on the collision energy based on the current measurement uncertainties. However, as expected due to the effect of the anisotropy of the collision region, peripheral collisions display a much higher $v_2$ than central collisions. Conversely, the triangular flow, which is driven predominantly by geometrical fluctuations in the overlap region, shows no significant centrality dependence.

\begin{figure}[h!]
\centering
\resizebox{.98\textwidth}{!}{%
\includegraphics[height=3cm]{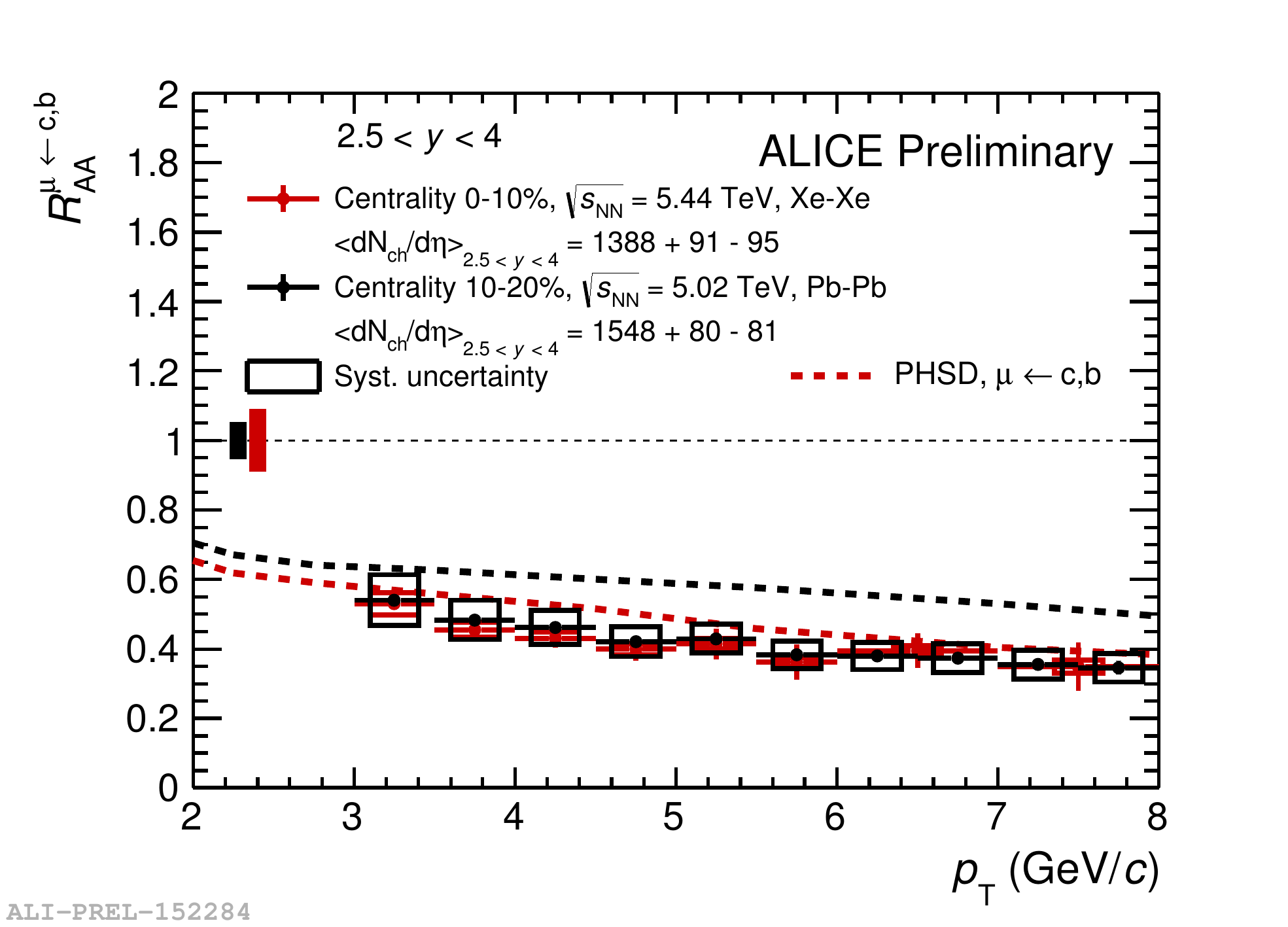}
\includegraphics[height=3cm]{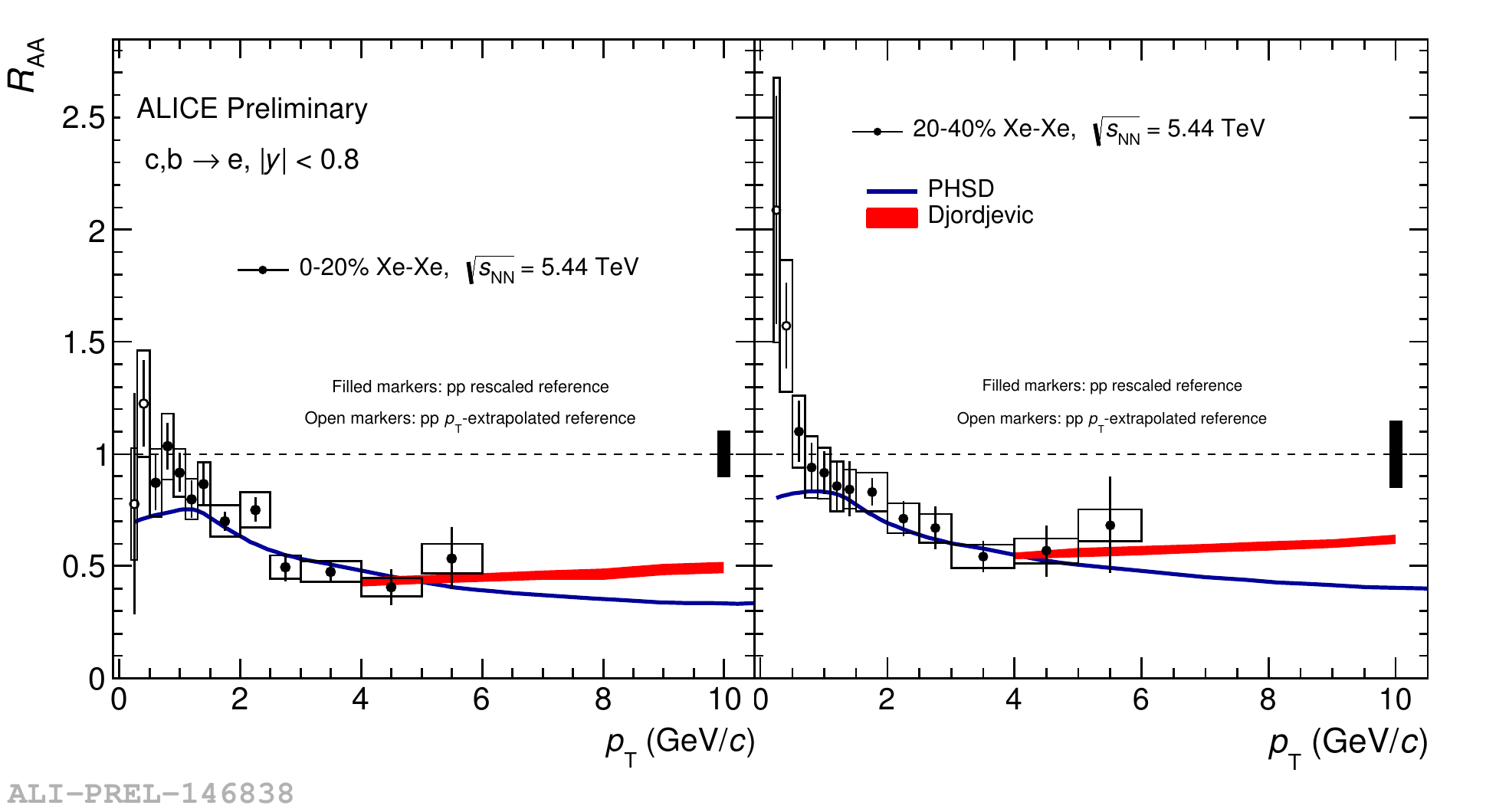}

}
\caption{Measurements of open heavy-flavour production in Xe--Xe and Pb--Pb collisions by the ALICE Collaboration, compared with theoretical model calculations. Left: Inclusive heavy-flavour decay muons in Xe--Xe and Pb--Pb collisions. Right: Inclusive heavy-flavour decay electrons in Xe--Xe collisions in two centrality classes.}
\label{fig:xexe}
\end{figure}

In 2017, a brief run was made of Xe--Xe collisions at the LHC. Due to the difference in system size, comparisons between Xe--Xe and Pb--Pb collisions allow the path length dependence of energy loss in the medium to be studied. This system has been studied by the ALICE Collaboration for inclusive heavy-flavour decay electrons at mid-rapidity and muons at forward rapidity. These results are shown in~\figref{fig:xexe}. The heavy-flavour decay electrons show a mild centrality dependence, with an \Raa that increases when going from central (0--20\%) to more peripheral (20--40\%) collisions. The results are described well by multiple theoretical models.
The heavy-flavour decay muons are shown in one centrality class (0--10\%) and compared with \PbPb collisions in 10--20\% centrality; the centralities for the comparison are chosen such that the charged-particle multiplicity in the collisions is similar. It can be seen that, although a difference in the \Raa is predicted by the PHSD model for the two collision systems, this prediction is not borne out by the data, where a clear agreement between $R_\mathrm{XeXe}$ and $R_\mathrm{PbPb}$ is seen. This result potentially implies that it is the particle multiplicity and not necessarily the system size itself that determines the degree of nuclear modification.

\section{Charmed baryon production}

\begin{figure}[h!]
\centering
\resizebox{.95\textwidth}{!}{%
\includegraphics[height=3cm]{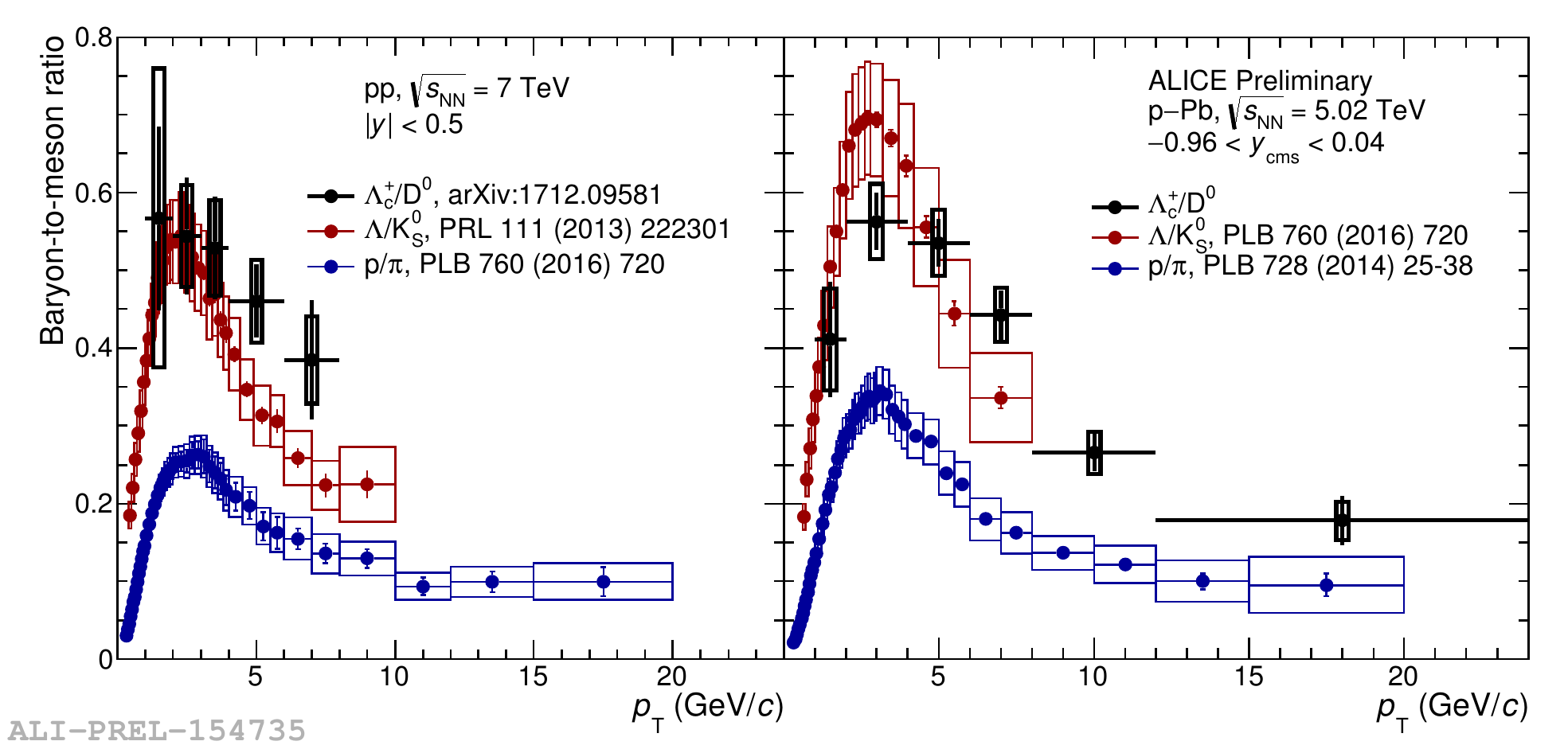}
\includegraphics[height=3.2cm, trim=0 0 280 200,clip]{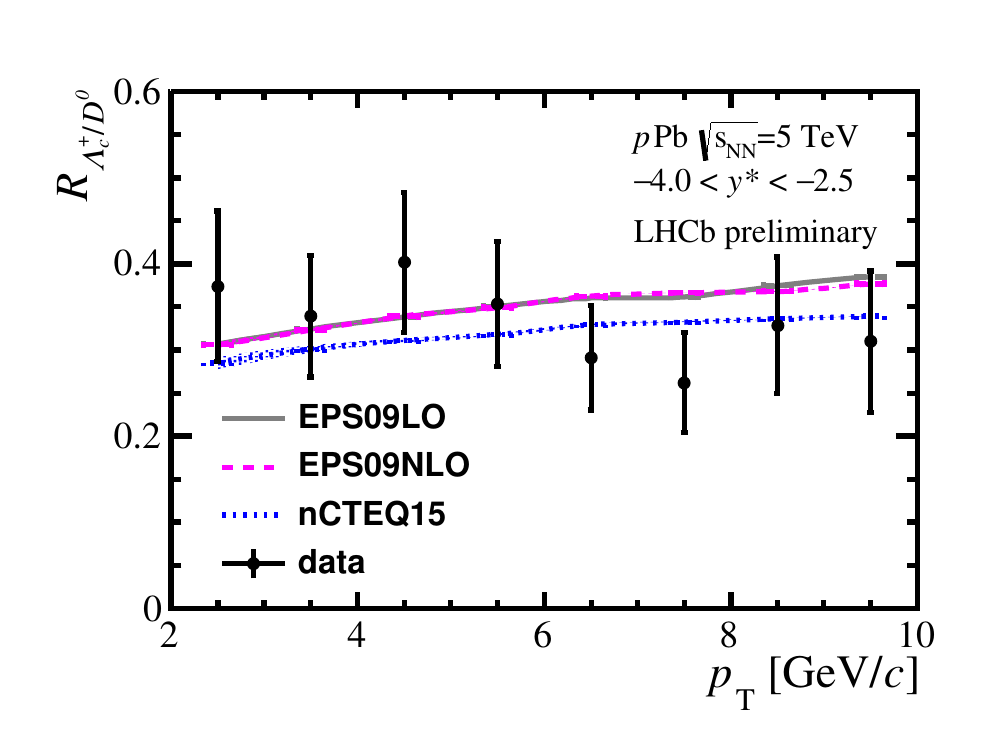}

}
\caption{Measurements of \Lc production in small systems at the LHC. Left: \Lc/\Dz production ratio in pp and \pPb collisions in ALICE, compared with light-flavour particles. Right: \Lc/\Dz production ratio in LHCb, compared with theoretical model calculations.}
\label{fig:lc1}
\end{figure}

Charmed baryons have currently been measured by both the ALICE and LHCb Collaborations at the LHC. The measurements of the \LcD baryon-to-meson ratio in pp and p--Pb collisions at mid-rapidity by ALICE and LHCb are shown in \figref{fig:lc1}.
The ALICE measurements show a consistent behaviour as a function of \pT for \Lc baryons at mid-rapidity in both pp and p--Pb collisions. The recent preliminary results from the Run 2 p--Pb data show a noticeable \pT dependence, with a decreasing ratio at high \pT. The results are compared with the equivalent quantities from the light-flavour sector, namely the p/$\pi$ and $\LKs$ ratios. Qualitatively, the \pT distributions of the baryon-to-meson ratios have a similar shape, and the magnitude of the \LcD ratio is similar to that of the \LKs ratio, potentially meaning that the hadronisation of strange and charmed baryons is due to similar processes in small systems. Meanwhile, the LHCb results from p--Pb collisions at forward rapidity show a lower \LcD ratio than the ALICE results, and a flat \pT distribution, implying a possible rapidity-dependent effect on charm baryon fragmentation in the absence of a medium. The data are described well by Monte Carlo models tuned to the existing LHCb results in proton--proton collisions.

\begin{figure}[h!]
\centering
\resizebox{.75\textwidth}{!}{%
\includegraphics[height=3cm]{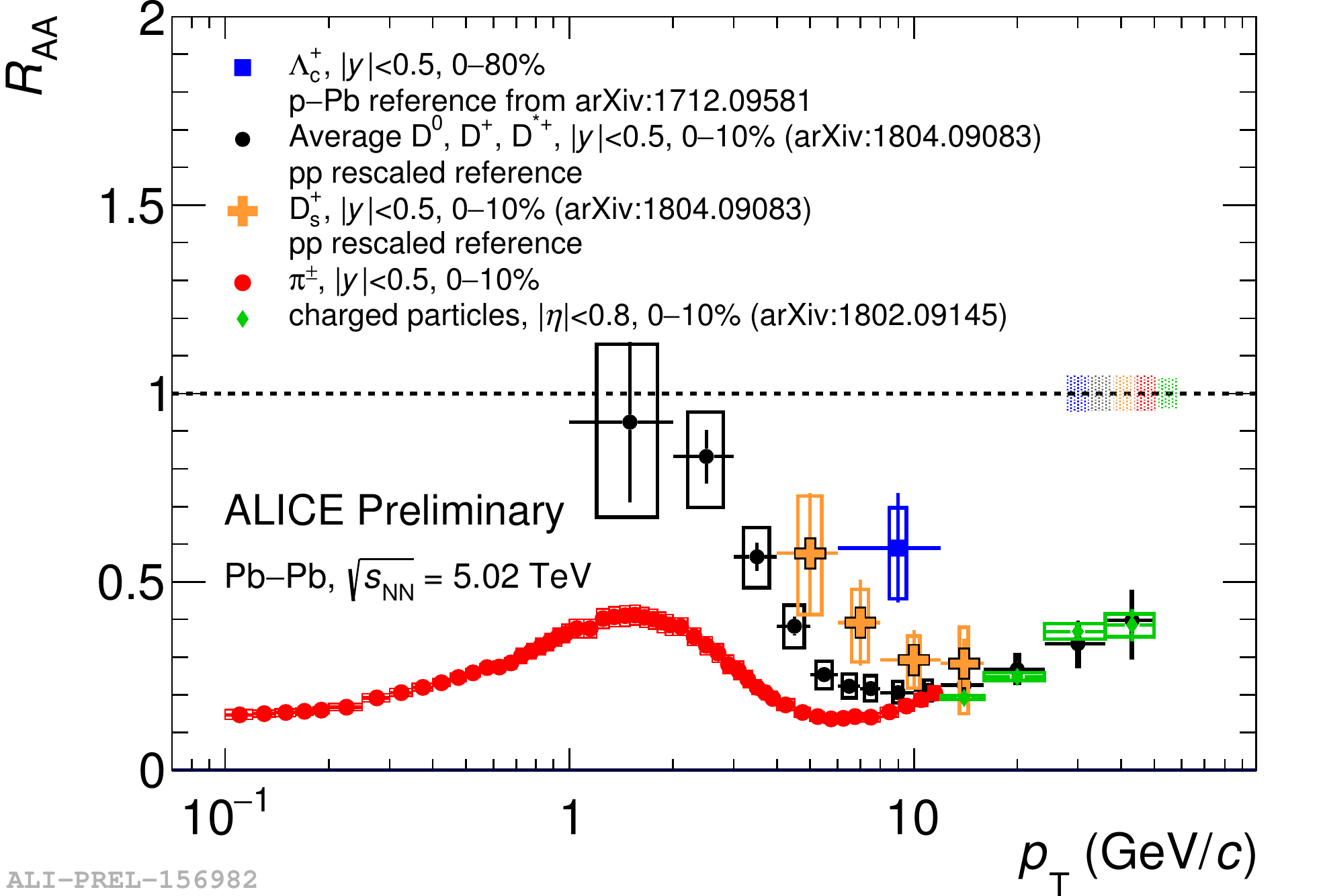}
\includegraphics[height=3cm]{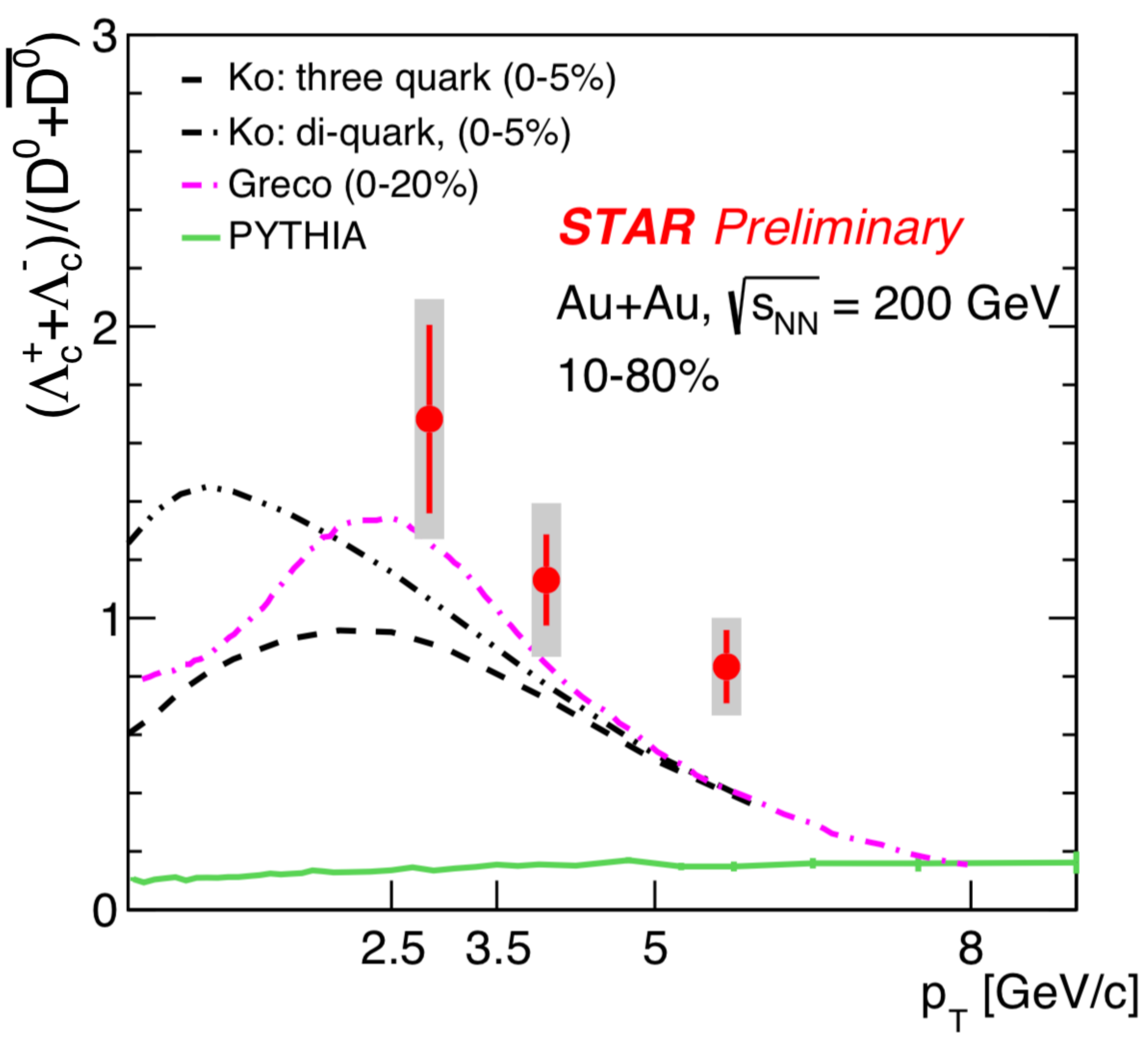}

}
\caption{Measurements of \Lc production in heavy-ion collisions. Left: \Raa of \Lc, strange and non-strange D mesons, and light charged particles, measured by ALICE in \PbPb collisions. Right: STAR measurement of \LcD ratio in Au--Au collisions at $\sqrt{s_\mathrm{NN}}=200\,\GeV$.}
\label{fig:lc2}
\end{figure}

\Figref{fig:lc2} shows the \Raa of \Lc baryons measured by ALICE, compared with strange D mesons, non-strange D mesons, and light-flavour charmed hadrons. A hint of a hierarchy in the \Raa can be seen in the \pT region where all particle species are measured; this seems to imply that \Lc baryons are less suppressed than strange D mesons, which are in turn less suppressed than non-strange D mesons and lighter hadrons. The \LcD ratio in \PbPb collisions is also increased above unity, implying a difference in the hadronisation mechanisms. A possible explanation for this is that quark recombination may play a significant role for charmed baryons in the medium. A similar effect is also seen by the STAR experiment at RHIC, where an increasing \LcD ratio is found in Au--Au collisions as a function of the collision centrality. The STAR results are described relatively well by models including quark coalescence, while currently the available models at LHC energies vastly underpredict the ALICE results. 

\section{Summary \& Outlook}

Many interesting properties of the production of heavy-flavour particles have been studied so far in heavy-ion collisions. The suppression of D and B mesons follows the hierarchy expected due to the mass dependence of quark energy loss. The $v_2$ of heavy particles in \pPb collisions hints at possible effects of collectivity in smaller systems at higher multiplicities, which was previously unexpected. Path length dependences of the energy loss have been studied by considering Xe--Xe and Pb--Pb collisions together, finding no significant dependence of the suppression of heavy-flavour particles on the system size. And the reduced suppression of \Dsubs, \Bsubs and \Lc compared with non-strange D and B mesons in heavy-ion collisions suggests the presence of unforeseen hadronisation mechanisms, for example the recombination of heavy quarks with lighter quarks within the medium.

Run 3 of the LHC will begin in 2021; over the course of the second Long Shutdown, significant upgrades will be made to both the experiments and to the accelerator itself, allowing a factor ${\sim}10$ increase in the recorded luminosity. This will lead to vast improvements in the precision of measurements for these rare particles, further constraining models, and will potentially allow new channels to be studied: for instance, the upgraded Inner Tracking System in the ALICE detector will allow the ALICE Collaboration to fully reconstruct B mesons in the central barrel, and increased statistics may make it possible to measure even rarer probes such as the $\Sigma_\mathrm{c}$ and $\Lambda_\mathrm{b}$ baryons in \PbPb collisions with high precision.

\section*{References}
\bibliography{beachproceedings}{}
\end{document}